\documentclass[%
reprint,
amsmath,amssymb,
aps,
prb,
]{revtex4-2}
\usepackage{graphicx}
\usepackage{dcolumn}
\usepackage{bm}
\usepackage{hyperref}

\begin{document}
\title{Plasmonic waveguides in two-dimensional materials: a quantum mechanical description using semiclassical techniques}

\author{T. M. Koskamp}
\email{tjacco.koskamp@ru.nl}
\author{M. I. Katsnelson}
\email{m.katsnelson@science.ru.nl}
\author{K. J. A. Reijnders}
\affiliation{
  Radboud University, Institute for Molecules and Materials, Heyendaalseweg 135, 6525 AJ Nijmegen, The Netherlands}
\date{\today}

\begin{abstract}
  Plasmons are likely to play an important role in integrated photonic ciruits, because they strongly interact with light and can be confined to subwavelength scales. These plasmons can be guided and controlled by plasmonic waveguides, which can be created by patterning different materials or by structuring the dielectric environment. We have constructed a semi-analytical theory to describe plasmonic waveguides, and, more generally, plasmons in spatially inhomogeneous systems. Our theory employs techniques from semiclassical analysis, and is therefore applicable when the electron wavelength is much smaller than the characteristic length scale of changes in the system parameters. We obtain an effective classical Hamiltonian that describes the dynamics of quantum plasmons, given by the Lindhard function with spatially varying parameters. Adding the wave-like character of the plasmons to the classical trajectories generated by this Hamiltonian, we find two different mechanisms for waveguiding. In the first case, a localized plasmonic state arises due to total internal reflection similar to photonic waveguides. The second mechanism relies on a varying dielectric environment, which locally modifies the screening of the electrons. Here, a quasi-localized state arises due to local changes in the amplitude of the plasmonic excitation. Our results provide a solid basis to understand previous numerical studies.
\end{abstract}

\maketitle

\section{Introduction}\label{sec:introduction}
Plasmons, quantized collective oscillations of conduction electrons in solids, may play an important role in the development of integrated photonic circuits, due to their strong coupling with light~\cite{Fitzgerald16,Tame13} and possible subwavelength confinement~\cite{Barnes03}. One way to control and guide plasmons is with the use of plasmonic waveguides, which can be created, for example, by patterning different materials or by structuring the dielectric environment. This can be done either invasively (e.g. cutting the active material~\cite{Grigorenko12,Basov16,Low17}), or non-invasively (e.g. changing the dielectric environment~\cite{Jiang21}), which makes plasmonic systems very versatile.  Recent experimental progress has made it possible to reach the quantum regime for plasmons~\cite{Scholl12,Fitzgerald16,Tame13}, which opens up a new world for plasmonic waveguides and is necessary for the practical application of plasmonic systems in general. However, in order to control (quantum) plasmons, one must, unavoidably, consider inhomogeneous systems.

The non-local character of the Coulomb interaction makes these inherently inhomogeneous systems difficult to describe in the quantum regime, both analytically and numerically. Numerically, this can be done by real space diagonalization within the framework of the random phase approximation~(RPA). These methods are often computationally intensive, as it amounts to the diagonalization of very large matrices and a very large basis set is needed. Numerical approaches for inhomogeneous quantum plasmonic systems are therefore often limited to small system sizes. With different methods, one can consider considerably larger systems~\cite{Yuan11,Yuan13,Jin15,Li23}, but this requires additional approximations, which are not always controllable.

Besides numerical approaches, it can be insightful and (computationally) effective to study analytical models for plasmonic systems. For classical systems, this can be effectively done using techniques from hydrodynamics, where one combines Maxwell's theory for electromagnetic waves in media and a Drude model for the behavior of the electrons in materials~\cite{Fitzgerald16,Krasavin08}. However, in the quantum regime for plasmons this is no longer applicable. A straightforward approach is to quantize the classical plasma, by discretizing the classical modes emerging from the Maxwell equations via the correspondence principle, which resembles a particle-in-a-box approach~\cite{Tame13,Fitzgerald16,Rice73}, but with a harmonic-oscillator like potential~\cite{Tame13,Archambault10,Huttner92}. Although accurate for specific systems, such methods do not incorporate the quantum character of the electron-electron interaction, and lose validity when the plasmon wavelength approaches the electron wavelength. 

In order to incorporate this interaction analytically, from a quantum perspective, a commonly used theory is the RPA. Because of the non-local character of the Coulomb interaction, one often turns to momentum space, making it difficult to work with spatially inhomogeneous systems like plasmonic waveguides. Recently published work~\cite{Reijnders22,Koskamp23} used the RPA, in combination with techniques from the semiclassical approximation, to obtain an (semi)analytical theory for plasmons in inhomogeneous systems, in the regime where the electron wavelength is much smaller than the characteristic length scale of the inhomogeneity. In Ref.~\cite{Koskamp23}, the authors considered two-dimensional~(2D) materials, using a simplifed model for the dielectric environment. The resulting theory was applied to scattering of plasmons on inhomogeneities. However, the techniques used in this approach can be extended to incorporate more realistic, and hence more complicated, dielectric environments.

In this article, we present a comprehensive theory for plasmonic waveguides in two-dimensional systems. We start by extending the theory discussed in Ref~\cite{Koskamp23} to 2D systems with an arbitrary out-of-plane dependence of the dielectric environment. We subsequently apply this general method to a commonly used model for layered structures described in e.g. Ref.~\cite{Jiang21,Emelyanenko08,Wehling11,Keldysh79}.In this model, the electrons are confined to a two-dimensional plane, which is surrounded by a layer with finite height $d$ and dielectric constant $\varepsilon_{\mathrm{M}}$. In turn, this layer is surrounded by two dielectrics with different dielectric constants.

By applying techniques from the semiclassical approximation to the RPA equations, we obtain an effective classical Hamiltonian for our 2D system. This effective classical Hamiltonian describes the dynamics of the quantum plasmons in phase space, and is given by the Lindhard function, but with spatially varying parameters. The classical plasmon trajectories generated by this Hamiltonian can be viewed as analogs of the rays in geometrical optics. By subsequently adding the wave-like character of the plasmons to these trajectories, we find two types of localized plasmonic states, which can potentially be used for waveguiding. 

The first type of localized state uses the principle of total internal reflection, analogous to photonic waveguides, in which a classically forbidden region emerges through the presence of a momentum along the direction of the waveguide~\cite{Born13}. We show that by varying the dielectric substrate, the electron density, or the effective height of the middle layer, it is possible to create plasmonic bound state. We systematically study the conditions for the appearance of these bound states, and compute their dispersion.

The second type of (quasi)localization relies on a varying dielectric substrate that locally modifies the screening of the electrons. In turn, locally increases or decreases the amplitude of the plasmonic excitation, thereby altering the induced electron density. We discuss how the localization of the plasmon depends on parameters like the substrate dielectric constant, the energy of the excited plasmon, and the momentum along the direction of the waveguide. These findings provide a solid basis to understand previous numerical results from Ref.~\cite{Jiang21}, but also show that it is possible to localize the plasmon in regions with higher screening.

This paper is set up in the following way. In Sec.~\ref{sec:GeneralDeriv}, we develop the theory for plasmons in 2D systems with an arbitrary out-of-plane dependence of the dielectric environment. We start with a brief review of the results obtained in Refs.~\cite{Reijnders22,Koskamp23} on how to apply techniques from the semiclassical approximation to the RPA. This introduces the concepts that are necessary to understand the rest of the paper. We subsequently obtain an effective classical Hamiltonian for quantum plasmons in generic 2D systems in Sec.~\ref{subsec:GeneralDerivPrincipalSymbol} . We also derive an expression for the Hartree potential (or induced potential). This induced plasmon potential mimics the wavefunction for particles subject to the Schr\"odinger equation. We show that the induced potential can be related to the induced electron density and to the electrostatic energy density, which has a physical interpretation as a probability density. Section~\ref{subsec:finitethickness} contains the application of this theory to the aforementioned dielectric model for layered materials, which results in the formulation of the effective classical Hamiltonian. In Sec.~\ref{sec:bound-states}, we extensively study bound states supported by this classical Hamiltonian. We elucidate their origin, and study how the spectrum depends on the spatial variation of different parameters. In Sec.~\ref{sec:Energy-Localized-states}, we discuss a second type of (quasi)localized states. We show how it depends on the screening from the substrate, as well as on the energy and momentum of the excited plasmon. Finally, we present our conclusions and a brief outlook in Sec.~\ref{sec:Conclusion}

\section{Effective description of plasmons in 2d systems with arbitrary dielectric environments}\label{sec:GeneralDeriv}

In Refs.~\cite{Reijnders22,Koskamp23}, a semi-analytical method to study plasmons in spatially inhomogeneous two-dimensional materials was developed. Its starting point is the equations of motion approach to the random phase approximation, which captures the electron-electron interaction through the Hartree potential. By applying techniques from semiclassical and microlocal analysis~\cite{Maslov81,Martinez02,Zworski12}, these equations were solved in the limit where the electron wavelength $\lambda_\mathrm{el}$ is much smaller than the characteristic scale $\ell$ of changes in the system parameters, i.e., the electron density and the background dielectric constant. Note that, while we refer to the dielectric constant as constant, it can still exhibit spatial variations as a function of $\mathbf{x}$. We assume the substrate material is locally characterized by a fixed static dielectric constant; however, in the presence of multiple materials, the dielectric properties can still vary spatially as a function of $\mathbf{x}$ to account for material transitions.

One of the main outcomes of this method is an effective classical Hamiltonian $\mathcal{H}_0(\mathbf{x},\mathbf{q},E)$, which captures the classical dynamics of the quantum plasmons. This quantity differs from the (Fourier transform of the) dielectric function $\overline{\varepsilon}(\mathbf{q},E)$, which is conventionally discussed when considering plasmons. Indeed, in the presence of translational invariance, the full dielectric function $\varepsilon(\mathbf{x},\mathbf{x}',E)$ can be written as $\varepsilon(\mathbf{x}-\mathbf{x}',E)$, which subsequently becomes $\overline{\varepsilon}(\mathbf{q},E)$ upon Fourier transform. In inhomogeneous systems, however, this reduction cannot be performed since translational invariance is broken.

In the limit $\lambda_\mathrm{el}/\ell \ll 1$, one may nevertheless say that we ``almost'' have translational invariance, since the external parameters, such as the Fermi momentum $p_\mathrm{F}$ and the background dielectric constant $\varepsilon_\mathrm{b}$, do not change significantly on the scale of the electron wavelength. This allows us to employ techniques from semiclassical analysis to obtain an effective classical Hamiltonian $\mathcal{H}_0(\mathbf{x},\mathbf{q})$, which depends on one position and one momentum variable. Although this Hamiltonian differs from the dielectric function $\overline{\varepsilon}(\mathbf{q},E)$, they turn out to be intimately related in the cases studied in Refs.~\cite{Reijnders22,Koskamp23}. More precisely, the effective classical Hamiltonian $\mathcal{H}_0(\mathbf{x},\mathbf{q},E)$ can be obtained from the Lindhard expression for $\overline{\varepsilon}(\mathbf{q},E)$ by replacing the parameters $p_\mathrm{F}$ and $\varepsilon_\mathrm{b}$ by the position-dependent functions $p_\mathrm{F}(\mathbf{x})$ and $\varepsilon_\mathrm{b}(\mathbf{x})$.

The next step of the method is to analyze the behavior of the classical Hamiltonian in phase space, in particular the trajectories that are generated by Hamilton's equations. This is completely analogous to the analysis of other Hamiltonians in classical mechanics, e.g., the Hamiltonian of the harmonic oscillator. The complete behavior of the plasmons can then be reconstructed by adding their wave-like character to the classical trajectories.

For two-dimensional systems, only the simplest model for the dielectric environment was studied in Ref.~\cite{Koskamp23}. As discussed in the introduction, more complicated models are often more suitable~\cite{Jiang21,Emelyanenko08,Wehling11}. In this section, we further extend the approach of Ref.~\cite{Koskamp23} to these more complicated models, and show that the aforementioned relation between the dielectric function and the effective classical Hamiltonian generally holds. To make this analysis self-contained, we first review the essential formulas from Refs~\cite{Reijnders22,Koskamp23} and sketch their physical context. For a more in-depth discussion of the underlying physical and mathematical concepts, including the notion of pseudodifferential operators and their symbols, we refer to the original articles.

\subsection{Review of the derivation of the effective classical Hamiltonian}
\label{subsec:reviewDerivClassHam}

We consider electrons that are confined to a two-dimensional plane $\mathbf{x}=(x,y)$, whose dynamics are governed by the single-electron Hamiltonian $\hat{H}_0$. Throughout this article, we assume that this single-electron Hamiltonian has the form $\hat{H}_0 = \hat{p}^2/2m + U(\mathbf{x})$. The potential $U(\mathbf{x})$ in this Hamiltonian can be related to a spatially varying electron density $n^{(0)}(\mathbf{x})$ using the Thomas-Fermi approximation~\cite{Vonsovsky89,Giuliani05,Lieb81}. A natural way to obtain a spatially varying $n^{(0)}(\mathbf{x})$ is to combine different materials.

In equilibrium, the electrons have a certain distribution, which can be described by the equilibrium density operator $\hat{\rho}_0$. When a weak perturbation is applied to the system, this equilibrium distribution is modified. In turn, this new electron distribution gives rise to a potential, which can be computed through the Poisson equation. In this way, a system of equations arises, which has to be solved self consistently~\cite{Vonsovsky89,Reijnders22,Koskamp23}. Within this framework, the plasmons are the self-sustained oscillations that remain after the external perturbation is switched off.

We therefore write the full Hamiltonian $\hat{H}$ of the system as the sum of the single-electron Hamiltonian $\hat{H}_0$ plus an additional Hartree potential $V_\mathrm{pl}$, that is, $\hat{H} = \hat{H}_0 + V_\mathrm{pl}$.
When the system is homogeneous, we can decompose this Hartee potential $V_\mathrm{pl}$ into Fourier modes. In our case, where the system is ``almost'' homogeneous on the scale of the electron wavelength, since $\lambda_\mathrm{el}/\ell \ll 1$, we can use the same Ansatz that is commonly used in the semiclassical approximation~\cite{Maslov81,Guillemin77}, namely
\begin{align}  \label{eq:scansatz2d}
  V_{\mathrm{pl}}(\mathbf{x}) = \varphi(\mathbf{x},\hbar)  e^{i S(\mathbf{x})/ \hbar} ,
\end{align}
where $S(\mathbf{x})$ is called the classical action, and where the amplitude $\varphi(\mathbf{x},\hbar)$ has a series expansion in powers of $\hbar$, that is,
\begin{align}  \label{eq:sc-amplitude-expansion}
  \varphi(\mathbf{x},\hbar) = \varphi_0(\mathbf{x}) + \hbar \varphi_1(\mathbf{x}) + \mathcal{O} (\hbar^2).
\end{align}
When this Ansatz is applied to one-dimensional problems in physics, one usually speaks of the Wentzel-Kramers-Brillouin~(WKB) approximation~\cite{Griffiths05}.

The induced electron density follows from the full Hamiltonian $\hat{H}$ by solving the Liouville-von Neumann equation of motion for the density operator.
For the semiclassical Ansatz~(\ref{eq:scansatz2d}), one can show that the induced density is given by~\cite{Reijnders22, Koskamp23}
\begin{equation}\label{eq:2delecdens}
  n(\mathbf{x}) = \big(\hat{\Pi} V_\mathrm{pl}\big) (\mathbf{x}) ,
\end{equation}
where, in contrast to the homogeneous case, the polarization $\hat{\Pi}$ is now a so-called pseudodifferential operator. These pseudodifferential operators can be viewed as generalizations of partial differential operators and are defined through their so-called symbols~\cite{Martinez02,Zworski12}, which are functions on classical phase space. Intuitively speaking, this relation can be understood as the correspondence between quantum mechanical operators and classical observables on phase space~\cite{Hall13}. Importantly, the symbols of pseudodifferential operators do not have to be polynomial in $\mathbf{p}$, as for partial differential operators, but can also have more complicated functional forms. A more detailed introduction to these operators can be found in Refs.~\cite{Martinez02,Zworski12}, and in Ref.~\cite{Reijnders22} in the context of the present formalism.

Because we use the semiclassical approximation, the symbol $\Pi$ of the operator $\hat{\Pi}$ naturally has an expansion in powers of $\hbar$, that is,
\begin{equation}  \label{eq:polarization-expansion-powers-h}
  \Pi(\mathbf{x},\mathbf{q},\hbar) = \Pi_0(\mathbf{x},\mathbf{q}) + \hbar \Pi_1(\mathbf{x},\mathbf{q}) + \mathcal{O}(\hbar^2) ,
\end{equation}
where $\Pi_0(\mathbf{x},\mathbf{q})$ is called the principal symbol. It is given by
\begin{align}  \label{eq:polarization-principal}
  \Pi_0 (\mathbf{x}, \mathbf{q}) \! = \! \frac{g_\mathrm{s}}{(2 \pi \hbar)^2} \! \int \! \frac{\rho_0\left(H_0(\mathbf{x},\mathbf{p})\right)-\rho_0\left(H_0(\mathbf{x},\mathbf{p}\!+\!\mathbf{q})\right)}{H_0(\mathbf{x},\mathbf{p})-H_0(\mathbf{x},\mathbf{p}\!+\!\mathbf{q}) + E} d \mathbf{p},
\end{align}
where $H_0(\mathbf{x},\mathbf{p}) = p^2/2m + U(\mathbf{x})$ is the symbol of the single-electron Hamiltonian. Moreover, $\rho_0(z)$ represents the Fermi-Dirac distribution.
It turns out that the integral in expression~(\ref{eq:polarization-principal}) can be evaluated in the same way as for a homogeneous system~\cite{Reijnders22,Koskamp23}. This yields the conventional Lindhard expression, but with parameters depending on the position, i.e., one replaces $p_\mathrm{F}$ by $p_\mathrm{F}(\mathbf{x})$.
As we previously mentioned, this allows us to interpret the principal symbol $\Pi_0(\mathbf{x},\mathbf{q})$ as the ``local'' polarization: it equals the polarization for a homogeneous electron gas, with parameters given by their values at the point $\mathbf{x}$.
In particular, Eq.~(\ref{eq:2delecdens}) implies that we can write the leading-order term of the induced density as
\begin{equation}
  n_0(\mathbf{x}) = \Pi_0\left(\mathbf{x},\frac{\partial S}{\partial \mathbf{x}}\right) \varphi_0(\mathbf{x}) e^{i S(\mathbf{x})/\hbar},
\end{equation}
to lowest order in the expansion parameter $\hbar$.

This induced density $n(\mathbf{x})$ in the layer at $z=0$ gives rise to an electrostatic potential $\Phi(\mathbf{x},z)$ through the Poisson equation. In two-dimensional materials, the strength of this electrostatic potential is strongly affected by the dielectric environment of the two-dimensional charge layer~\cite{Jiang21,Koskamp23}, since this environment screens the Coulomb interaction between charges at different positions. This screening becomes more important when the distance between two charges increases, so it is especially important in the limit of small $\mathbf{q}$. We denote the dielectric environment of the surrounding media by $\varepsilon(\mathbf{x},z)$, explicitly indicating that it can vary in both the in-plane and out-of-plane directions.
We can then write the Poisson equation as
\begin{equation}  \label{eq:Poisson}
  \left\langle \nabla , \varepsilon(\mathbf{x},z) \nabla \right\rangle V(\mathbf{x},z) = - 4 \pi e^2 n(\mathbf{x},z),
\end{equation}
where $V(\mathbf{x},z)$ is related to the electrostatic potential by $V=-e\Phi$, with $e$ the elementary charge, and the induced density equals $n(\mathbf{x},z) =  n(\mathbf{x}) \delta (z)$. In Eq.~(\ref{eq:Poisson}), $\nabla = \left(\partial/ \partial \mathbf{x} , \,\partial / \partial z\right)$ denotes the three-dimensional gradient and $\langle a , b \rangle$ denotes the three-dimensional Cartesian inner product between the vectors $a$ and $b$. Note that we do not consider external electric fields in the Poisson equation, because we are interested in plasmons, which are self-sustained collective oscillations.

It is important to note that the Hartree potential $V_\mathrm{pl}(\mathbf{x})$ in the Hamiltonian is caused by the potential $V(\mathbf{x},z)$ obtained from Eq.~(\ref{eq:Poisson}). In order to ensure that our set of equations is self-consistent, we therefore impose the additional condition~\cite{Koskamp23}
\begin{equation}  \label{eq:condition-self-consistency}
  V(\mathbf{x},z=0) = V_\mathrm{pl}(\mathbf{x}) .
\end{equation}
By combining Eqs.~(\ref{eq:2delecdens}),~(\ref{eq:Poisson}) and~(\ref{eq:condition-self-consistency}), we can obtain expressions for the action $S(\mathbf{x})$ and the amplitude $\varphi(\mathbf{x})$ in the asymptotic solution~(\ref{eq:scansatz2d}), as well as construct an effective classical Hamiltonian for the two-dimensional quantum plasmons. 

In order to construct the Hamiltonian, we need an additional physical ingredient. As noted in Ref.~\cite{Koskamp23}, the in-plane variables $\mathbf{x}$ can be regarded as ``slow'' variables, since the system parameters do not change significantly on the scale of the electron wavelength. The out-of-plane variables $z$ can, instead, be considered ``fast'' variables, since the system parameters can change significantly on the scale of the electron wavelength. Because of this partition in fast and slow variables, we can perform an adiabatic separation of the in-plane and out-of-plane degrees of freedom in the potential $V(\mathbf{x},z)$, similar to the Born-Oppenheimer approximation. In the original formulation of the Born-Oppenheimer approximation, one employs an instantaneous eigenfunction that depends on the variables $\mathbf{x}$ and $z$. As explained in Ref.~\cite{Belov06}, this formulation does not suffice when one deals with a rapidly oscillating exponent~(\ref{eq:scansatz2d}). Instead, one has to consider a slightly more complicated form for the potential $V(\mathbf{x},z)$, namely~\cite{Koskamp23}
\begin{align}  \label{eq:opsepvar}
  V(\mathbf{x},z) = (\hat{\Gamma} V_\mathrm{pl})(\mathbf{x},z) ,
\end{align}
where $\hat{\Gamma}$ is a pseudodifferential operator. This Ansatz yields a generalized Born-Oppenheimer approximation, in which the instantaneous eigenfunction is replaced by an operator. Similar to Eq.~(\ref{eq:polarization-expansion-powers-h}), the symbol $\Gamma(\mathbf{x},\mathbf{q},z,\hbar)$ of the operator $\hat{\Gamma}$ can be expanded in powers of $\hbar$, yielding a principal symbol $\Gamma_0(\mathbf{x},\mathbf{q},\hbar)$ and a subprincipal symbol $\Gamma_1(\mathbf{x},\mathbf{q},\hbar)$.
Comparing these symbols to the instantaneous eigenfunctions in the original Born-Oppenheimer approach, we may loosely say that the generalized form~(\ref{eq:opsepvar}) adds the momentum variable to the original Ansatz.

Inserting the Ansatz~(\ref{eq:opsepvar}) into Eq.~(\ref{eq:Poisson}) and taking Eq.~(\ref{eq:2delecdens}) into account, one can convert the Poisson equation into two ordinary differential equations for the principal and subprincipal symbols $\Gamma_0$ and $\Gamma_1$, respectively, of the operator $\hat{\Gamma}$. This construction makes extensive use of the calculus for pseudodifferential operators, discussed earlier in this section, and is performed in Ref.~\cite{Koskamp23} order by order in $\hbar$.
In the end, one finds that the principal symbol $\Gamma_0$ satisfies the ordinary differential equation
\begin{equation} \label{eq:opsepvarhbar0}
  F_0\left(\mathbf{x},\mathbf{q},z,\frac{\partial}{\partial z}\right) \Gamma_0(\mathbf{x},\mathbf{q},z)  = - 4 \pi e^2 \delta(z) \Pi_0 (\mathbf{x},\mathbf{q}),
\end{equation}
where $F_0$ is given by
\begin{equation} \label{eq:leading-ordersymbolODE}
  F_0\left(\mathbf{x},\mathbf{q},z,\frac{\partial}{\partial z}\right) = -\frac{|\mathbf{q}|^2}{\hbar^2} \varepsilon(\mathbf{x},z)  + \frac{\partial}{\partial z} \left(\varepsilon(\mathbf{x},z)\frac{\partial}{\partial z}\right) .
\end{equation}
Solving Eq.~(\ref{eq:opsepvarhbar0}), one finds an explicit form for $\Gamma_0$.

We are now ready to construct the effective classical Hamiltonian.
Inserting the Ansatz~(\ref{eq:scansatz2d}) and Eq.~(\ref{eq:opsepvar}) into the self-consistency condition~(\ref{eq:condition-self-consistency}), one finds the secular equation
$\mathcal{H}_0(\mathbf{x},\partial S/\partial\mathbf{x}) \varphi_0(\mathbf{x}) \exp(i S/\hbar)=0$,
where~\cite{Koskamp23}
\begin{equation}\label{eq:defEffClasHam}
  \mathcal{H}_0 (\mathbf{x},\mathbf{q}) = 1 - \Gamma_0 (\mathbf{x},\mathbf{q},z=0) .
\end{equation}
This secular equation is equivalent to the Hamilton-Jacobi equation $\mathcal{H}_0(\mathbf{x},\partial S/\partial\mathbf{x})=0$ for the action $S(\mathbf{x})$. This implies that $\mathcal{H}_0$ can be interpreted as the effective classical Hamiltonian that describes the dynamics of quantum plasmons. 

We can thus summarize this subsection by saying that one obtains an effective classical Hamiltonian for two-dimensional quantum plasmons by solving the ordinary differential equation~(\ref{eq:opsepvarhbar0}) and inserting the solution into Eq.~(\ref{eq:defEffClasHam}). In the next subsection, we study the structure of this ordinary differential equation in detail, and construct the effective classical Hamiltonian $\mathcal{H}_0 (\mathbf{x},\mathbf{q})$ for a general class of models for $\varepsilon(\mathbf{x},z)$.

\subsection{General construction of the effective classical Hamiltonian}
\label{subsec:GeneralDerivPrincipalSymbol}

In Ref.~\cite{Koskamp23}, the construction of the effective classical Hamiltonian $\mathcal{H}_0$ was performed using the simplest model for the dielectric environment, in which the background dielectric environment does not depend on $z$, and is described by $\varepsilon_\mathrm{A}(\mathbf{x})$ and  $\varepsilon_\mathrm{B}(\mathbf{x})$ above and below the layer, respectively. This leads to an effective dielectric function in the plasmon dispersion, which arises due to screening effects from the surrounding environment. In this simplest approximation, this function is given by the average of the dielectric properties of the involved media. However, as discussed in the introduction, more complicated models for the dielectric environment are often necessary to more accurately describe the screening effects. In homogeneous systems, such advanced models typically yield more accurate predictions of three-dimensional screening effects, as they account for non-local interactions. Consequently, the effective dielectric function in these cases is no longer a simple averaged quantity, but instead exhibits a $\mathbf{q}$-dependence~\cite{Jiang21,Emelyanenko08,Wehling11}. We would therefore like to extend the formalism developed in Ref.~\cite{Koskamp23} to include more complicated models for the dielectric environment. 

According to Eq.~(\ref{eq:defEffClasHam}), we can obtain the effective classical Hamiltonian $\mathcal{H}_0$ for a given model of the dielectric environment $\varepsilon(\mathbf{x},z)$ by determining $\Gamma_0$, which is the solution of Eq.~(\ref{eq:opsepvarhbar0}). We can construct a general solution to this inhomogeneous differential equation using the method of variation of parameters~\cite{Korn00}. With this method, we can express $\Gamma_0$ in terms of the fundamental solutions of the homogeneous differential equation. This leads to a general solution for $\Gamma_0$ that is independent of the specific model for $\varepsilon(\mathbf{x},z)$, but nevertheless gives us many important physical insights, as we will see shortly. We subsequently consider a specific model for the dielectric environment $\varepsilon(\mathbf{x},z)$ in Sec.~\ref{subsec:finitethickness}.

Let us therefore first consider the fundamental solutions of the homogeneous differential equation, i.e.,
\begin{equation}\label{eq:homogeneousODE}
  \left( \frac{\partial}{\partial z} \left(\varepsilon(\mathbf{x},z)\frac{\partial}{\partial z}\right) -\frac{1}{\hbar^2} \varepsilon(\mathbf{x},z) |\mathbf{q}|^2 \right) w_{\mathrm{H}}(z) = 0 .
\end{equation}
This equation has two solutions, which we denote by $w_1$ and $w_2$, and whose specific form depends on the form of $\varepsilon(\mathbf{x},z)$.
In order to make some progress, we assert that $\varepsilon(\mathbf{x},z)$ goes to a constant as $z \to \pm \infty$. This condition is physically very intuitive, and does not limit the practical applicability of our theory, since it does not dictate the precise shape of $\varepsilon(\mathbf{x},z)$. Because of our assertion, the fundamental solutions $w_{1,2}$ are asymptotically equivalent to a linear combination of the functions $\exp(\pm |\mathbf{q}| z / \hbar)$ in the limit $z \to \pm \infty$, as follows from Eq.~(\ref{eq:homogeneousODE}).
By making use of the freedom in our choice of the asymptotic solutions, we can then construct them in such a way that $w_1 \to 0$ as $z \to \infty$, and $w_2 \to 0$ as $z \to -\infty$.

The solution of the inhomogeneous differential equation~(\ref{eq:opsepvarhbar0}) can subsequently be expressed as~\cite{Korn00}
\begin{equation} \label{eq:totalsolinhomoeq}
  \Gamma_{0}(z) = c_1 (z) w_1(z) + c_2(z) w_2(z) .
\end{equation}
We find the functions $c_1(z)$ and $c_2(z)$ by inserting Eq.~(\ref{eq:totalsolinhomoeq}) into the inhomogeneous differential equation. Using standard arguments from the method of variation of parameters~\cite{Korn00}, we then find a set of two differential equations, which can be combined into
\begin{equation}\label{eq:diffeqforcz}
  W^{T}\begin{pmatrix}
    \frac{\mathrm{d}c_1}{\mathrm{d} z} \\
    \frac{\mathrm{d}c_2}{\mathrm{d} z}
  \end{pmatrix}
  =
  \begin{pmatrix}
    0 \\
    f(z)
  \end{pmatrix},
\end{equation}
where $W$ resembles the Wronskian matrix, and is given by
\begin{equation}
  W = \begin{pmatrix}
    w_1 & \varepsilon \frac{\mathrm{d} w_1}{\mathrm{d} z}\\
    w_2 & \varepsilon \frac{\mathrm{d} w_2}{\mathrm{d} z}
  \end{pmatrix}.
\end{equation}
The function $f(z) = - 4 \pi e^2 \Pi_0 \delta(z)$ corresponds to the inhomogeneous term in Eq.~(\ref{eq:opsepvarhbar0}).
Because of the homogeneous differential equation~(\ref{eq:homogeneousODE}), the derivative of the determinant $\det(W)$ with respect to $z$ vanishes, which means that $\det(W)$ is constant.
Moreover, $\det(W)$ is nonzero when the fundamental solutions $w_{1,2}$ are linearly independent, meaning that the matrix $W$ is invertible.

On physical grounds, we require that the potential $V(\mathbf{x},z)$, see Eqs.~(\ref{eq:Poisson}) and~(\ref{eq:opsepvar}), goes to zero as $z\to\pm\infty$. In turn, this implies that $\Gamma_0 \to 0$ as $z\to\pm\infty$. Because of the way in which we constructed $w_{1,2}(z)$, this condition implies that $c_1 \to 0$ as $z \to -\infty$ and $c_2 \to 0$ as $z \to \infty$.
We may therefore write
\begin{equation}
\begin{aligned}
  \int_{-\infty}^z \frac{\mathrm{d}c_1}{\mathrm{d} z'} d z' &= c_1(z) - c_1(-\infty) = c_1 (z), \\
  \int^{\infty}_z \frac{\mathrm{d}c_2}{\mathrm{d} z'} d z' &= c_2(\infty) - c_2(z) = -c_2 (z).
\end{aligned}
\label{eq:dc12dz-var-constants}
\end{equation}
Rewriting the solution~(\ref{eq:totalsolinhomoeq}) for $\Gamma_0(z)$ using Eq.~(\ref{eq:dc12dz-var-constants}), and inserting expressions for the derivatives obtained from Eq.~(\ref{eq:diffeqforcz}), we obtain, cf. Ref.~\cite{Korn00},
\begin{align} \label{eq:generalsol-Gamma0}
  \Gamma_{0}(z) 
  &= \left(\int_{-\infty}^z \frac{\mathrm{d}c_1}{\mathrm{d} z'} d z'\right) w_1(z) - \left( \int^{\infty}_z \frac{\mathrm{d}c_2}{\mathrm{d} z'} d z' \right) w_2(z) \nonumber \\
  &= \int_{-\infty}^{\infty} g(z,z') f(z') d z',
\end{align}
where 
\begin{equation}\label{eq:generalsol-Green}
  g(z,z') = \begin{cases}
    - \frac{1}{\mathrm{det} W} w_2(z') w_1 (z),  &-\infty < z'< z\\
    - \frac{1}{\mathrm{det} W} w_1(z') w_2 (z),  &z < z' < \infty
  \end{cases}.
\end{equation}
The function $g(z,z')$ is, of course, a Green's function.

Inserting our expression for the inhomogeneous term $f(z) = - 4 \pi e^2 \Pi_0 \delta(z)$, we now easily find that
\begin{equation}\label{eq:definition-principal-symbol}
  \Gamma_0 (z) = 
  \begin{cases}
    \frac{2 \pi e^2 \hbar }{\varepsilon_{\mathrm{eff}} |\mathbf{q}|} \Pi_0 \frac{w_1(z)}{w_1 (0)},   &z>0\\
    \frac{2 \pi e^2 \hbar }{\varepsilon_{\mathrm{eff}} |\mathbf{q}|}\Pi_0  \frac{w_2(z)}{w_2(0)},  &z<0
  \end{cases},
\end{equation}
where we defined an effective dielectric function $\varepsilon_\mathrm{eff}$ as
\begin{align}\label{eq:definition-effective-dielectricconstant}
  \varepsilon_{\mathrm{eff}} = \frac{\hbar}{2 |\mathbf{q}|}  \frac{\mathrm{det}W}{w_1(0) w_2(0)},
\end{align}
which captures the three-dimensional screening from the environment. Using the definition of $W$, this expression can also be written as
\begin{align}\label{eq:definition-effective-dielectricconstant2}
  \varepsilon_{\mathrm{eff}} = \frac{\hbar}{2 |\mathbf{q}|} \varepsilon (0)\left( \frac{1}{w_2(0)}\frac{\mathrm{d}w_2}{\mathrm{d} z}(0) -  \frac{1}{w_1(0)}\frac{\mathrm{d}w_1}{\mathrm{d} z} (0) \right).
\end{align}
This dielectric function should not be confused with the longitudinal Lindhard (dielectric) function, which describes the dynamical screening of the electric field by electrons in the 2D electron gas. Instead, that screening is incorporated into our theory through the effective classical Hamiltonian, which we obtain by substituting the expression for $\Gamma_0(z)$ into the self-consistency condition Eq.~(\ref{eq:condition-self-consistency}). This yields  
\begin{equation}\label{eq:effective-classical-Hamiltonian}
  \mathcal{H}_0 (\mathbf{x},\mathbf{q}) = 1 - \frac{2 \pi e^2 \hbar}{\varepsilon_\mathrm{eff}(\mathbf{x},\mathbf{q}) |\mathbf{q}|} \Pi_0 (\mathbf{x},\mathbf{q}),
\end{equation}
where we now explicitly indicated the dependence on $\mathbf{x}$ and $\mathbf{q}$.

Let us take a step back and reflect on the physical implications of this result. Equation~(\ref{eq:effective-classical-Hamiltonian}) shows that we can capture the effect of any dielectric environment by a single effective dielectric function $\varepsilon_\mathrm{eff}(\mathbf{x},\mathbf{q})$ in the effective classical Hamiltonian. This is an important generalization of Ref.~\cite{Koskamp23}, where this effective dielectric function was only computed for the simplest model of the dielectric environment. Here, we have instead considered an arbitrary shape of $\varepsilon(\mathbf{x},z)$, only imposing the condition that it becomes constant as $z\to\pm\infty$. Moreover, we have expressed the effective dielectric function in terms of the value of the fundamental solutions to the homogeneous differential equation~(\ref{eq:homogeneousODE}) at the point $z=0$, which makes our result easily applicable to different dielectric environments.

We note that the specific model that is commonly used for layered structures with an encapsulated thin film with finite thickness $d$ also leads to an effective dielectric function $\varepsilon_\mathrm{eff}$ in the Lindhard function~\cite{Jiang21,Emelyanenko08,Wehling11}. As we already discussed, the effective classical Hamiltonian can be viewed as, and takes the form of, the Lindhard function with position-dependent parameters. The result~(\ref{eq:effective-classical-Hamiltonian}) may therefore seem somewhat straightforward. However, we would like to stress two key findings. First of all, it does not depend on the specific model that is used. In some sense, one may therefore say that expression~(\ref{eq:effective-classical-Hamiltonian}), and more specifically Eq.~(\ref{eq:definition-effective-dielectricconstant2}) also extends the known results for the homogeneous case. Second, our result is not just valid for the homogeneous case, but also, and especially, for the inhomogeneous case, as long as $\lambda_\mathrm{el}/\ell \ll 1$ as discussed in the introduction. In Sec.~\ref{subsec:finitethickness}, we discuss the specific model for layered structures with a finite thickness $d$ and show that expression~(\ref{eq:definition-effective-dielectricconstant}) indeed leads to the conventional expression for $\varepsilon_\mathrm{eff}$, but with position-dependent parameters.

As discussed in detail in Ref.~\cite{Koskamp23}, the effective classical Hamiltonian~(\ref{eq:effective-classical-Hamiltonian}) can be used to analyze the behavior of the quantum plasmon in classical phase space $(\mathbf{x},\mathbf{q})$. One can compute the classical trajectories using Hamilton's equations, which are equivalent to the Hamilton-Jacobi equation  $\mathcal{H}_0 (\mathbf{x},\partial S/ \partial \mathbf{x}) = 0$. The classical action $S(\mathbf{x})$ in this equation is related to the momentum $\mathbf{q}$ by $\mathbf{q} = \partial S/ \partial \mathbf{x} $ and enters the Hartree potential $V_\mathrm{pl}$, cf. Eq.~(\ref{eq:scansatz2d}). In general, open trajectories in phase space correspond to plasmon scattering, which was analyzed in Ref.~\cite{Koskamp23}. Closed trajectories, on the other hand, correspond to bound states and plasmon quantization. We will consider these bound states in detail in Sec.~\ref{sec:bound-states}, in which we also analyze the classical Hamiltonian~(\ref{eq:effective-classical-Hamiltonian}).

\subsection{Review of the derivation of the amplitude $\varphi_0$}
\label{subsec:reviewDerivAmplitude}

In the previous subsections, we discussed the derivation of the classical Hamiltonian $\mathcal{H}_0$. While performing this derivation, we also found the defining equation for the classical action $S(\mathbf{x})$ in the Ansatz~(\ref{eq:scansatz2d}) for the Hartree potential, namely the Hamilton-Jacobi equation.
In this subsection, we consider the amplitude $\varphi_0$ in the Ansatz~(\ref{eq:scansatz2d}). We briefly review its derivation based on Ref.~\cite{Koskamp23}, to which we refer for further details.

In the final step of the derivation of the classical Hamiltonian in Sec.~\ref{subsec:reviewDerivClassHam}, we inserted both the Ansatz~(\ref{eq:scansatz2d}) and expression~(\ref{eq:opsepvar}) into the self-consistency condition~(\ref{eq:condition-self-consistency}). Gathering the terms of order $\hbar^0$ in the result, one finds that the amplitude $\varphi_0$ drops out. In order to obtain an equation for this amplitude, we should therefore consider the terms of order $\hbar^1$ in the self-consistency condition. On the right-hand side of this condition, one has
\begin{align} 
  & \hspace*{-0.25cm} (\hat{\Gamma} V_{\mathrm{pl}}) (\mathbf{x},z) = e^{ i S(\mathbf{x})/ \hbar} \Bigg(\Gamma_0  \varphi_0 + \hbar \Gamma_0  \varphi_1 -i \hbar \left\langle\frac{\partial \Gamma_0}{\partial \mathbf{q}},\frac{\partial \varphi_0}{\partial \mathbf{x}}\right\rangle    \Bigg.  \nonumber \\
  & \hspace*{0.25cm} + \Bigg. \hbar\Gamma_1  \varphi_0
  - \frac{i \hbar}{2} \sum_{j,k} \frac{\partial^2 \Gamma_0}{\partial q_j \partial q_k} \frac{\partial^2 S}{\partial x_j \partial x_k} \varphi_0\Bigg) + \mathcal{O}(\hbar^2),
  \label{eq:fullpotential}
\end{align}
where $\Gamma_0$ and $\Gamma_1$ are to be evaluated at the point $(\mathbf{x},\partial S/\partial \mathbf{x},z)$. Although we did not determine the quantity $\Gamma_1$ yet, we continue with this expression. We come back to the defining equations for $\Gamma_1$, also called the subprincipal symbol, at the end of this subsection.

Gathering all terms of order $\hbar^1$ in the self-consistency condition~(\ref{eq:condition-self-consistency}) with the help of Eq.~(\ref{eq:fullpotential}), and using our previous definition~(\ref{eq:defEffClasHam}) of the effective classical Hamiltonian, we find that~\cite{Koskamp23}
\begin{align}\label{eq:transporteq}
  &\mathcal{H}_1\left(\mathbf{x},\frac{\partial S}{\partial \mathbf{x}}\right) \varphi_0 - i \left\langle\frac{\partial \mathcal{H}_0}{\partial \mathbf{q}}\left(\mathbf{x},\frac{\partial S}{\partial \mathbf{x}}\right), \frac{\partial \varphi_0}{\partial \mathbf{x}} \right\rangle \nonumber \\
  &- \frac{i}{2} \sum_{j,k} \frac{\partial^2 \mathcal{H}_0}{\partial q_j \partial q_k}\left(\mathbf{x},\frac{\partial S}{\partial \mathbf{x}}\right) \frac{\partial^2 S}{\partial x_k \partial x_j} \varphi_0 = 0 ,
\end{align}
where we defined $\mathcal{H}_1 \equiv -\Gamma_1(\mathbf{x},\partial S/\partial \mathbf{x},z = 0)$. This equation is known as the transport equation in the literature~\cite{Maslov81,Koskamp23}.

The transport equation can be solved using standard semiclassical techniques, which can be found in, e.g. Refs.~\cite{Maslov81,Koskamp23}. We briefly review its main steps. The construction starts from the classical trajectories of the effective classical Hamiltonian~(\ref{eq:defEffClasHam}), which are the solutions of Hamilton's equations. These trajectories can be labeled by the time $\tau$ along the trajectory and a parameter $\alpha$, which reflects the initial conditions and distinguishes different trajectories. In a scattering problem, the parameter $\alpha$ parameterizes the initial wavefront, see e.g. Ref.~\cite{Koskamp23} for a complete example. One subsequently introduces the Jacobian $J(\mathbf{x}) = \det(\partial \mathbf{x} / \partial (\tau,\alpha))$, and defines the quantity $A_0(\mathbf{x})$ by $A_0 (\mathbf{x})\equiv \varphi_0 (\mathbf{x}) \sqrt{J (\mathbf{x})}$. Making use of Hamilton's equations and the Liouville formula~\cite{Maslov81,Reijnders22,Koskamp23}, one then obtains an ordinary differential equation for $A_0(\mathbf{x})$ along the trajectories of the dynamical system, namely
\begin{align}
  \frac{\mathrm{d} A_0}{\mathrm{d} t} + i\bigg(\mathcal{H}_1 + \frac{i}{2} \sum_j \frac{\partial^2 \mathcal{H}_0}{\partial x_j \partial q_j}\bigg)A_0 = 0,
\end{align}
which has the straightforward solution
\begin{align}\label{eq:amplitudesol}
  A_0 (\mathbf{x}) = A_0^0 \exp\bigg(-i \int_0^t  \mathcal{H}_1 + \frac{i}{2} \sum_j \frac{\partial^2 \mathcal{H}_0}{\partial x_j \partial q_j} d t'\bigg) .
\end{align}
The integral in this expression is to be performed along the trajectories of the Hamiltonian system.
Note that, strictly speaking, the above derivation is only valid when there is a one-to-one mapping of the trajectories onto configuration space $\mathbf{x}$, that is, in the absence of turning points. However, as argued in Refs.~\cite{Reijnders22,Koskamp23}, one can incorporate the presence of turning points in the description by introducing the Maslov index. We will come back to this point in Sec.~\ref{sec:bound-states}.

Equation~(\ref{eq:amplitudesol}) gives us the solution for the amplitude $\varphi_0(\mathbf{x}) = A_0 (\mathbf{x})/\sqrt{J (\mathbf{x})}$ in expression~(\ref{eq:scansatz2d}).
However, as we already mentioned in the beginning of this subsection, this solution contains the so far undetermined quantity $\mathcal{H}_1(\mathbf{x},\mathbf{q}) = -\Gamma_1(\mathbf{x},\mathbf{q},z=0)$. In the final part of this subsection, we therefore consider the equations that define this quantity, before solving them in the next subsection.

As in Sec.~\ref{subsec:reviewDerivClassHam}, we insert Eq.~(\ref{eq:opsepvar}) into Eq.~(\ref{eq:Poisson}) and take Eq.~(\ref{eq:2delecdens}) into account. Making use of the calculus for pseudodifferential operators, as discussed in detail in Ref.~\cite{Koskamp23}, we find that $\Gamma_1$ is determined by the ordinary differential equation
\begin{align} 
  &F_1\left(\mathbf{x},\mathbf{q},z\right) \Gamma_0(\mathbf{x},\mathbf{q},z) + F_0\bigg(\mathbf{x},\mathbf{q},z,\frac{\partial}{\partial z}\bigg) \Gamma_1(\mathbf{x},\mathbf{q},z) \nonumber \\
  &\hspace*{1.3cm} - i \left\langle\frac{\partial F_0}{\partial \mathbf{q}} \bigg(\mathbf{x},\mathbf{q},z,\frac{\partial}{\partial z}\bigg),\frac{\partial \Gamma_0}{\partial \mathbf{x}} (\mathbf{x},\mathbf{q},z)\right\rangle \nonumber \\
  &\hspace*{3.0cm} = - 4\pi e^2 \delta(z) \Pi_1 (\mathbf{x},\mathbf{q}),
  \label{eq:opsepvarhbar1}
\end{align}
where $F_1$ is defined by 
\begin{multline} \label{eq:subleading-ordersymbolODE}
  F_1(\mathbf{x},\mathbf{q},z) 
    = \frac{i}{\hbar^2} \left\langle\mathbf{q},\frac{\partial \varepsilon}{\partial \mathbf{x}}(\mathbf{x},z)\right\rangle \\
    = -\frac{i}{2} \sum_{j} \frac{\partial^2 F_0}{\partial q_j \partial x_j}(\mathbf{x},\mathbf{q},z) ,
\end{multline}
where we made use of Eq.~(\ref{eq:leading-ordersymbolODE}) in the final equality.

The quantity $\Pi_1$ in Eq.~(\ref{eq:opsepvarhbar1}) is the so-called subprincipal symbol of the polarization operator~\cite{Koskamp23}. Since $\hat{\Pi}$ is a Hermitian operator, the following relation between the principal and the subprincipal symbol always holds~\cite{Martinez02,Reijnders22}:
\begin{equation}  \label{eq:polarization-subprincipal-principal-general}
  \text{Im} \, \Pi_1(\mathbf{x},\mathbf{q}) = - \frac{1}{2}\sum_j \frac{\partial \Pi_0}{\partial q_j \partial x_j} (\mathbf{x},\mathbf{q}) .
\end{equation}
In the specific case that we consider in this paper, where the electron Hamiltonian takes the form $\hat{H}=\hat{p}^2/2m + U(\mathbf{x})$, we have an even simpler relation~\cite{Reijnders22}, namely
\begin{equation}  \label{eq:polarization-subprincipal}
  \Pi_1(\mathbf{x},\mathbf{q}) = -\frac{i}{2}\sum_j \frac{\partial \Pi_0}{\partial q_j \partial x_j} (\mathbf{x},\mathbf{q}) .
\end{equation}
In the remainder of this paper, we will assume that Eq.~(\ref{eq:polarization-subprincipal}) holds. At the same time, we will briefly explore the consequences of the relation~(\ref{eq:polarization-subprincipal-principal-general}), which is more general and also holds for other electron Hamiltonians. For a more detailed discussion on the background of $\Pi_1$, we refer to Ref.~\cite{Reijnders22}.

\subsection{General construction of the potential and interpretation through energy density} \label{subsec:GeneralDerivAmplitude}

In the previous subsection, we reviewed the derivation of the amplitude $\varphi_0$ and found that this amplitude contains the quantity $\mathcal{H}_1=-\Gamma_1(z=0)$. In turn, the quantity $\Gamma_1$ is the solution of the ordinary differential equation~(\ref{eq:opsepvarhbar1}).
In this subsection, we construct an expression for $\Gamma_1$ using variation of parameters, in the same way as we constructed an expression for $\Gamma_0$ in Sec.~\ref{subsec:GeneralDerivPrincipalSymbol}. As before, this solution generalizes the discussion in Ref.~\cite{Koskamp23}, where $\Gamma_1$ was only constructed for the simplest case, in which the dielectric environment does not depend on $z$. Here, we consider an arbitrary $\varepsilon(\mathbf{x},z)$, only assuming that it becomes constant as $z\to\pm\infty$.
With our expression for $\Gamma_1$, we subsequently compute the amplitude $\varphi_0$ according to the theory reviewed in the previous subsection. We finally show that the electromagnetic energy density computed with the potential $V(\mathbf{x},z)$ can be interpreted as a probability density in the semiclassical sense, again generalizing the results from Ref.~\cite{Koskamp23}.

Let us therefore consider the differential equation~(\ref{eq:opsepvarhbar1}). We first note that the only difference between the differential equations for $\Gamma_0$ and $\Gamma_1$, Eqs.~(\ref{eq:opsepvarhbar0}) and~(\ref{eq:opsepvarhbar1}), is the inhomogeneous term. The homogeneous differential equations are exactly the same. We can therefore use our previous fundamental solutions $w_1(z)$ and $w_2(z)$ to construct an expression for $\Gamma_1$. Let us denote the inhomogeneous term in Eq.~(\ref{eq:opsepvarhbar1}) by
\begin{equation}  \label{eq:def-f1}
  f_1(z) = f_{1s}(z) -4 \pi e^2 \Pi_1 \delta(z) ,
\end{equation}
where we omitted the dependence on $(\mathbf{x},\mathbf{q})$ and
\begin{equation}  \label{eq:def-f1s}
  f_{1s}(z) = -F_1(z) \Gamma_0(z)  + i \left\langle\frac{\partial F_0}{\partial \mathbf{q}} \bigg(z,\frac{\partial}{\partial z}\bigg),\frac{\partial \Gamma_0}{\partial \mathbf{x}} (z)\right\rangle .
\end{equation}
Drawing on the discussion in Sec.~\ref{subsec:GeneralDerivPrincipalSymbol}, we can then write
\begin{equation} \label{eq:generalsol-Gamma1}
  \Gamma_{1}(z) = \int_{-\infty}^{\infty} g(z,z') f_1(z') d z',
\end{equation}
cf. Eq.~(\ref{eq:generalsol-Gamma0}), where $g(z,z')$ is given by Eq.~(\ref{eq:generalsol-Green}).

We now note that we are not interested in the full expression for $\Gamma_1$, but only in its value at $z=0$, since this value enters the amplitude~(\ref{eq:amplitudesol}) through $\mathcal{H}_1=-\Gamma_1(z=0)$.
We can compute this value by performing the integration in Eq.~(\ref{eq:generalsol-Gamma1}). This integration is straightforward for the part with the delta function, but more involved for $f_{1s}$, which also contains our expression~(\ref{eq:definition-principal-symbol}) for $\Gamma_0$. 
Nevertheless, the integral can be calculated explicitly, using integration by parts, as we show in Appendix~\ref{subapp:amplitude}.

The final result can be cast in the form
\begin{align}
  \mathcal{H}_1 + \frac{i}{2} \sum_j \frac{\partial^2 \mathcal{H}_0}{\partial x_j \partial q_j}  &=  -\frac{2 \pi e^2 \hbar}{\varepsilon_{\mathrm{eff}}|\mathbf{q}| } \left(\Pi_1 + \frac{i}{2}\sum_j \frac{\partial^2 \Pi_0}{\partial x_j \partial q_j} \right) \nonumber\\ 
  & \qquad- \frac{i}{2}\left\{\ln \varepsilon_{\mathrm{eff}}|\mathbf{q}| , \mathcal{H}_0 \right\},\label{eq:AmplitudeExponent}
\end{align}
where $\{ a , \mathcal{H}_0 \}$ denotes the Poisson bracket, defined in Eq.~(\ref{eq:def-Poisson-bracket}), of $a$ and the effective Hamiltonian $\mathcal{H}_0$.
Because of Hamilton's equations, this Poisson bracket can be written as a total derivative with respect to time~\cite{Goldstein02,Arnold89}. This greatly simplifies the integration in the expression~(\ref{eq:amplitudesol}) for the amplitude, cf. the discussion in Ref.~\cite{Koskamp23}, and leads to
\begin{equation}\label{eq:Amplitude}
  A_0 (\mathbf{x}) =  \frac{A_0^0}{\sqrt{\varepsilon_{\mathrm{eff}}(\mathbf{x},\partial S / \partial \mathbf{x}) |\partial S / \partial \mathbf{x}|}} e^{i \Phi_\mathrm{B} (\mathbf{x})} ,
\end{equation}
where $\mathbf{q}$ has become $\partial S / \partial \mathbf{x}$ in the denominator because we integrate along the trajectories of the Hamiltonian system.
The quantity $\Phi_\mathrm{B}$ in Eq.~(\ref{eq:Amplitude}) is the Berry phase, defined by
\begin{equation}  \label{eq:Berry-phase-definition}
  \Phi_\mathrm{B}(\mathbf{x}) = \int_0^t \frac{2 \pi e^2 \hbar}{\sqrt{\varepsilon_{\mathrm{eff}} |\mathbf{q}|}} \left(\Pi_1 + \frac{i}{2} \sum_j \frac{\partial^2 \Pi_0}{\partial q_j \partial x_j}\right) dt.
\end{equation}
Because of Eq.~(\ref{eq:polarization-subprincipal}), the Berry phase is zero for a parabolic electron Hamiltonian $H_0 = \hat{p}^2/(2m) + U(\mathbf{x})$. However, for more complicated electron Hamiltonians the term in parentheses need not be zero. Nevertheless, Eq.~(\ref{eq:polarization-subprincipal-principal-general}) indicates that Eq.~(\ref{eq:Berry-phase-definition}) is purely real and therefore indeed a phase.

With the help of Eqs.~(\ref{eq:opsepvar}) and~(\ref{eq:scansatz2d}), we now obtain the leading-order term of the full potential $V(\mathbf{x},z) = \hat{\Gamma} V_{\mathrm{pl}}$, namely 
\begin{align}
  V(\mathbf{x},z) &= \varphi_0(\mathbf{x}) \Gamma_0\left(\mathbf{x},\frac{\partial S}{\partial \mathbf{x}},z\right) e^{i S(\mathbf{x})/\hbar}\\
  &= \frac{A_0^0 e^{i \Phi_\mathrm{B}(\mathbf{x})}e^{i S(\mathbf{x})/\hbar}}{\sqrt{J(\mathbf{x})\varepsilon_{\mathrm{eff}}(\mathbf{x},\partial S / \partial \mathbf{x}) |\partial S / \partial \mathbf{x}|}} \Gamma_0\left(\mathbf{x},\frac{\partial S}{\partial \mathbf{x}},z\right) ,
  \label{eq:full-potential}
\end{align}
where $\Gamma_0$ is given by Eq.~(\ref{eq:definition-principal-symbol}).
As discussed in Sec.~\ref{subsec:GeneralDerivPrincipalSymbol}, the action $S(\mathbf{x})$ in this expression can be calculated from the Hamilton-Jacobi equation, or, equivalently, from integrating Hamilton's equations.
We have thus found an expression for the full potential in real space, which is not only valid at $z=0$, where the charged layer is situated, but also outside of it.
It is valid for arbitrary $\varepsilon(\mathbf{x},z)$, provided that this function goes to a constant as $z\to\pm\infty$. Note that this assumption ensures that the potential decays exponentially for $z\to\pm\infty$.

Let us take a closer look at the physical interpretation to Eq.~(\ref{eq:full-potential}). We first note that Eq.~(\ref{eq:full-potential}) differs from semiclassical approximations to the solution of the Schr\"odinger equation, which have the form $\psi = A_0^0 \exp(i S/\hbar)/\sqrt{J}$. One can give a physical interpretation to these expressions by considering the probability density $|\psi|^2\propto 1/|J|$. The factor $1/|J|$ ensures that the probability density is invariant under a coordinate transformation, see the discussion in Refs.~\cite{Guillemin77,Zworski12,Koskamp23}. Clearly, Eq.~(\ref{eq:full-potential}) cannot be interpreted in the same way, since $V$ is  a different quantity: it is not a solution to the Schr\"odinger equation. Instead, it is related to an electrostatic potential $\Phi$ through $V=-e\Phi$. We may therefore consider the energy density stored in the electromagnetic field coming from the potential $V(\mathbf{x},z)$.

In Ref.~\cite{Koskamp23}, this energy density was computed for the case where the background dielectric does not depend on $z$. It was shown that the electromagnetic energy density, when integrated along $z$, is proportional to $1/|J|$, that is, has the same mathematical structure as $|\psi|^2$ for the Schr\"odinger equation. This means that the integrated energy density indeed behaves as a density, and provides an additional physical interpretation of the potential $V(\mathbf{x},z)$. Let us check whether the same conclusion holds for an arbitrary $\varepsilon(\mathbf{x},z)$.

Following the derivation in Ref.~\cite{Koskamp23}, the energy density for $z\neq 0$ is given by
\begin{equation}\label{eq:totalenergydensity}
  \mathcal{U}(\mathbf{x}, z) = \frac{\varepsilon (\mathbf{x}, z)}{16 \pi e^2 } \left|\nabla V(\mathbf{x},z)\right|^2 ,
\end{equation}
where the gradient is three-dimensional, meaning that one should take the derivative with respect to both $\mathbf{x}$ and $z$.
In Appendix~\ref{subapp:energydensity}, we compute the leading-order term of the energy density~(\ref{eq:totalenergydensity}) for the potential~(\ref{eq:full-potential}) and show that it satisfies
\begin{equation}\label{eq:inplaneenergydensity}
  \mathcal{U}_\mathrm{I}(\mathbf{x}) = \int_{-\infty}^\infty \mathcal{U}(\mathbf{x},z) dz  =\frac{1}{8 \pi e^2 } \frac{|A_0^0|^2}{|J(\mathbf{x})|}.
\end{equation}
The integrated energy density is therefore exactly the same as Eq.~(81) in Ref.~\cite{Koskamp23}. However, we have now proved this formula for an arbitrary dielectric $\varepsilon(\mathbf{x},z)$. This shows that $\mathcal{U}_\mathrm{I}$ indeed has the mathematical structure of a density~\cite{Guillemin77,Zworski12} and thereby provides an additional physical interpretation of the potential $V(\mathbf{x},z)$.

\subsection{Plasmons in layered structures with an effective height $d$} \label{subsec:finitethickness}
The theory discussed in the previous subsections is valid for arbitrary static dielectric environment $\varepsilon (\mathbf{x},z)$, provided that it goes to a constant for $z\to\pm\infty$, and in that sense completely general. In the remainder of this section, we apply this general theory to a specific model for the dielectric environment. More specifically, we discuss layered structures where a thin film with dielectric constant $\varepsilon_\mathrm{M}$ and finite effective height $d$ is encapsulated by two semi-infinite layers above and below, with dielectric constants $\varepsilon_\mathrm{A}$ and $\varepsilon_\mathrm{B}$, respectively. This model leads to an effective description for plasmons that captures the non-local Coulomb screening from the substrate~\cite{Jiang21,Wehling11,Rosner16}, and was shown to be accurate for describing electrons in for example TMDC's~\cite{Steinke20,Wehling11}.

In this subsection, we derive the effective classical Hamiltonian for this model, and show that $\varepsilon_\mathrm{eff}$ is given by the expression known from the literature~\cite{Jiang21,Emelyanenko08,Wehling11}, but with position-dependent parameters. Our aim is not only to show consistency, i.e. to show that our formalism leads to the same results for the homogeneous case, but more importantly to extend this well-known result to the inhomogeneous case. In other words, as long as $\lambda_\mathrm{el}/\ell \ll 1$, we can obtain the effective classical Hamiltonian $\mathcal{H}_0$ for the inhomogeneous case from the Lindhard function for the homogeneous case by replacing the parameters by position-dependent variables.

From the derivation in Sec.~\ref{subsec:GeneralDerivPrincipalSymbol}, we see that the principal symbol $\Gamma_0(z)$ leads to the effective classical Hamiltonian $\mathcal{H}_0$. In turn, $\Gamma_0(z)$ is determined by the fundamental solutions $w_1$ and $w_2$ to the homogeneous differential equation~(\ref{eq:homogeneousODE}). With the principal symbol $\Gamma_0(z)$, we can also compute the full leading-order potential $V(\mathbf{x},z)$ and the energy density $\mathcal{U}_\mathrm{I}(\mathbf{x})$, see the discussion in Sec.~\ref{subsec:GeneralDerivAmplitude}.

We consider a system where the electrons are bound to the 2D plane at $z=0$, that is, $n(\mathbf{x},z) = n(\mathbf{x})\delta(z)$, and where the total dielectric environment as function of $z$ takes the form
\begin{equation}\label{eq:varepsilon}
  \varepsilon(\mathbf{x},z) = \varepsilon_i(\mathbf{x}) =
  \begin{cases}
    \varepsilon_\mathrm{A}(\mathbf{x})  , & z>d/2 \\
    \varepsilon_\mathrm{M}(\mathbf{x})  , & d/2>z>-d/2\\
    \varepsilon_\mathrm{B}(\mathbf{x})  , & z<-d/2
  \end{cases},
\end{equation}
where the transition at $\pm d$ is instantaneous and behaves like a step function. We call the $\varepsilon_i$ dielectric constants, since we assume that each layer has a locally well-defined static dielectric constant, which does not depend on momentum or energy. However, these dielectric constants can still vary spatially across $\mathbf{x}$, for instance because of a variation of different materials in each layer.
Since we assume that the spatial scale $\ell$ of variations in the $\varepsilon_i$ is much larger than the electron wavelength, these dielectric constants are well-defined on the scale of the electron wavelength.

The two independent solutions to the homogeneous differential equation have the same form, namely
\begin{equation}\label{eq:generalsolution}
  w_i(z) =
  \begin{cases}
    c_{\mathrm{A}^+} e^{|\mathbf{q}| z / \hbar} + c_{\mathrm{A}^-} e^{-|\mathbf{q}| z / \hbar} 	, & z>d/2 \\
    c_{\mathrm{M}^+} e^{|\mathbf{q}| z / \hbar} + c_{\mathrm{M}^-} e^{-|\mathbf{q}| z / \hbar}   , & d/2>z>-d/2\\
    c_{\mathrm{B}^+} e^{|\mathbf{q}| z / \hbar} + c_{\mathrm{B}^-} e^{-|\mathbf{q}| z / \hbar} 	, & z<-d/2
  \end{cases},
\end{equation}
where the constants are determined from the boundary conditions. We obtain these conditions directly from the differential equation~(\ref{eq:homogeneousODE}): both $w_i$ and $\varepsilon(\mathbf{x},z) d w_i/d z$ have to be continuous at the boundaries $\pm d$.

We note that, in principle, the fundamental solutions are not uniquely defined, since any linear combination of two fundamental solutions is again a fundamental solution. By demanding that $w_1$ ($w_2$) decays for $z \to \infty$ ($z \to -\infty$), we determine it uniquely up to normalization. In what follows, we construct $w_1$ explicitly. Because of the symmetry, the second solution $w_2$ is then easily found by changing $z$ to $-z$, and interchanging $\varepsilon_{\mathrm{A}}$ and $\varepsilon_{\mathrm{B}}$.

Since $w_1$ decays as $z\to \infty$, we have $c_{\mathrm{A}^+} = 0$ in Eq.~(\ref{eq:generalsolution}). The remaining constants are determined by the boundary conditions. We obtain
\begin{widetext}
\begin{equation}
  w_1(z) =
  \frac{1}{\left(1 + \tilde{\varepsilon}_\mathrm{B}e^{ - 2 |\mathbf{q}| d / \hbar}\right)\left(1+ \tilde{\varepsilon}_\mathrm{A}e^{ - 2 |\mathbf{q}| d / \hbar}\right)}
   \begin{cases}
    \left(1+ \tilde{\varepsilon}_\mathrm{A}\right)e^{-|\mathbf{q}| z / \hbar} 	, & z\!>\!d/2 \\
    \tilde{\varepsilon}_\mathrm{A} e^{ -2 |\mathbf{q}| d / \hbar} e^{|\mathbf{q}| z / \hbar} + e^{-|\mathbf{q}| z / \hbar}   , & d/2\!>\!z\!>\!-d/2\\
    \frac{\tilde{\varepsilon}_\mathrm{A} e^{- 2 |\mathbf{q}| d / \hbar} + \tilde{\varepsilon}_\mathrm{B}e^{2 |\mathbf{q}| d / \hbar}}{1+\tilde{\varepsilon}_\mathrm{B}}e^{|\mathbf{q}| z / \hbar} + \frac{1+ \tilde{\varepsilon}_\mathrm{A}\tilde{\varepsilon}_\mathrm{B} e^{-4|\mathbf{q}| d / \hbar} }{1+\tilde{\varepsilon}_\mathrm{B}} e^{-|\mathbf{q}| z / \hbar} 	, & z\!<\!-d/2 
  \end{cases}, \label{eq:fundamentalsolution1}
\end{equation}
\end{widetext}
where $\tilde{\varepsilon}_i = (\varepsilon_{\mathrm{M}}-\varepsilon_i )/( \varepsilon_{\mathrm{M}}+\varepsilon_i ) $.
Note that the normalization factor, which is arbitrary, has no physical consequences. Looking at our expression~(\ref{eq:definition-effective-dielectricconstant2}) for $\varepsilon_{\mathrm{eff}}$, we clearly see that it is divided out.

Following the general theory discussed in Sec.~\ref{subsec:GeneralDerivPrincipalSymbol}, we find the principal symbol $\Gamma_0$ from Eq.~(\ref{eq:definition-principal-symbol}), which yields 
\begin{equation}
  \Gamma_0 (z)\! =\! \frac{2 \pi e^2 \hbar }{\varepsilon_{\mathrm{eff}} |\mathbf{q}|} \Pi_0 \begin{cases}
    \frac{1+ \tilde{\varepsilon}_\mathrm{A}}{1  + \tilde{\varepsilon}_\mathrm{A} e^{-2\frac{d}{\hbar}|\mathbf{q}|}}e ^{-\frac{z}{\hbar}|\mathbf{q}|}, &\!z\!>\!d/2\\
    \frac{e^{-\frac{z}{\hbar}|\mathbf{q}|} + \tilde{\varepsilon}_\mathrm{A}  e^{-2\frac{d}{\hbar}|\mathbf{q}|} e^{\frac{z}{\hbar}|\mathbf{q}|}}{1  + \tilde{\varepsilon}_\mathrm{A} e^{-2\frac{d}{\hbar}|\mathbf{q}|} }, &\!<\!z\!<\!d/2\\
    \frac{\tilde{\varepsilon}_\mathrm{B}  e^{-2\frac{d}{\hbar}|\mathbf{q}|} e^{-\frac{z}{\hbar}|\mathbf{q}|} + e^{\frac{z}{\hbar}|\mathbf{q}|}}{1  + \tilde{\varepsilon}_\mathrm{B} e^{-2\frac{d}{\hbar}|\mathbf{q}|} }, &\!-d/2\!<\!z\!<\!0 \\
    \frac{1+ \tilde{\varepsilon}_\mathrm{B}}{1  + \tilde{\varepsilon}_\mathrm{B} e^{-2\frac{d}{\hbar}|\mathbf{q}|}}e^{-\frac{z}{\hbar}|\mathbf{q}|},    &\!z\!<\!-d/2 \\
  \end{cases},
  \label{eq:Gamma0-finite-thickness}
\end{equation}
where
\begin{equation} \label{eq:effective-dielectric-constant-finite-thickness}
  \varepsilon_{\mathrm{eff}} 
    = \varepsilon_\mathrm{M}  \frac{1  - \tilde{\varepsilon}_\mathrm{A}\tilde{\varepsilon}_\mathrm{B}e^{ - 2 |\mathbf{q}| d / \hbar}}{1 + \left(\tilde{\varepsilon}_\mathrm{A}+\tilde{\varepsilon}_\mathrm{B}\right)e^{ -  |\mathbf{q}| d / \hbar} + \tilde{\varepsilon}_\mathrm{A}\tilde{\varepsilon}_\mathrm{B}e^{ - 2 |\mathbf{q}| d / \hbar}},
\end{equation}
as defined by Eq.~(\ref{eq:definition-effective-dielectricconstant}). 
Having computed the effective dielectric constant $\varepsilon_{\mathrm{eff}}$, we immediately obtain the effective classical Hamiltonian~(\ref{eq:effective-classical-Hamiltonian}) and the full potential, given by Eq.~(\ref{eq:full-potential}).

Equation~(\ref{eq:effective-dielectric-constant-finite-thickness}) indeed corresponds to the well-known expression from the literature~\cite{Jiang21,Emelyanenko08,Wehling11}, but with position-dependent parameters. We have therefore extended this result to the inhomogeneous case. As previously discussed, the effective classical Hamiltonian can be viewed as the analog of the Lindhard function, with the parameters corresponding to their values at a given point $\mathbf{x}$. In the next section, we will study this Hamiltonian in more detail, in order to get a physical understanding of this model. More specifically, we will look at in-plane variations of various parameters, and show that these variations allow for the existence of plasmonic bound states.

\section{Plasmon localization through bound states}\label{sec:bound-states}
In this section, we analyze the effective classical Hamiltonian~(\ref{eq:effective-classical-Hamiltonian}) at zero temperature and show that it allows for the formation of bound states. These states arise when a classically allowed region lies in between two regions where plasmon propagation is classically forbidden. In a plasmonic waveguide, these classically forbidden regions emerge through the presence of a momentum along the direction of propagation in the waveguide.

Throughout this section, we consider the model discussed in Sec.~\ref{subsec:finitethickness}, where the substrate layer above (A) and below (B) have the same background (b) dielectric constant, that is, $\varepsilon_\mathrm{A} (\mathbf{x}) = \varepsilon_\mathrm{B}(\mathbf{x}) = \varepsilon_\mathrm{b}(\mathbf{x})$ in Eq.~(\ref{eq:definition-effective-dielectricconstant}). We use the term dielectric constant to indicate that this quantity is static and does not depend on the out-of-plane coordinate $z$. However, it still depends on the in-plane coordinates $x$. At the same time, the length scale of variations in this direction is large compared to the electron wavelength, which justifies the term dielectric constant. 

In Sec.~\ref{subsec:GeneralAnalysisBoundStates}, we analyze the classical Hamiltonian for this model. We show that plasmonic bound states can arise in a waveguide geometry by spatially varying the dielectric constant, but only when the momentum along the propagation direction of the waveguide is non-zero. In Sec.~\ref{subsec:NumImplBoundStates}, we subsequently implement this setup in Wolfram Mathematica~\cite{Mathematica}, and compute the bound-state spectrum. We not only consider variations in the dielectric constant, but also in the electron density $n^{(0)}$ and the effective height $d$, introduced in the previous section.

\subsection{General analysis of the effective classical Hamiltonian: formation of bound states}\label{subsec:GeneralAnalysisBoundStates}
We first demonstrate that the effective classical Hamiltonian supports plasmonic bound states. We study a system in which the parameters vary only in the $x$-direction, while the system is translationally invariant in the $y$-direction. This implies that $\mathcal{H}_0(x,\mathbf{q},E)$ does not explicitly depend on $y$, which means  $d q_y / d \tau = - \partial \mathcal{H}_0 / \partial y$ because of Hamilton's equations. In other words, $q_y$ is conserved and thus serves as a good quantum number. We call the $y$-direction the propagation direction of the waveguide. 

As previously mentioned, we can interpret the classical Hamiltonian as a spatially varying analog of the conventional Lindhard function, where the parameters take their local values at position $x$. A plasmon mode can exist at a given position $x_i$ if there is a real wavevector $\mathbf{q}$ satisfying $\mathcal{H}_0 (x_i,\mathbf{q},E) = 0$, for a given energy $E$. From the relation $\mathcal{H}_0 (x,\mathbf{q},E) = 0$, we can compute the local plasmon dispersion $E(x;\mathbf{q})$, treating $x$ as a parameter. When there is no real wavevector $\mathbf{q}$ satisfying this relation, we speak of a classically forbidden region.

At first sight, it may seem strange that a 2D plasmonic waveguide exhibits classically forbidden regions, since the 2D plasmon spectrum is gapless as $\mathbf{q} \to 0$. This is in contrast to the 3D plasmon spectrum, which has a cutoff at the plasma frequency $\omega_\mathrm{p}$. This energy gap, which depends on the dielectric constant and the electron density, results in classically forbidden regions which can give rise to bound states~\cite{Reijnders22}. Although the 2D plasmon spectrum is gapless as a function of $|\mathbf{q}|$, the presence of a finite momentum $q_y$ along the propagation direction creates an effective gap for propagation in the $x$-direction, as we discuss shortly. This mechanism is  analogous to total internal reflection in photonic waveguides, where a critical angle determines whether a photon is completely reflected~\cite{Born13}. From this analogy, it is clear that a non-zero momentum in the direction of propagation is required.

\begin{figure}[t]
  \centering
  \includegraphics[width=0.95\linewidth]{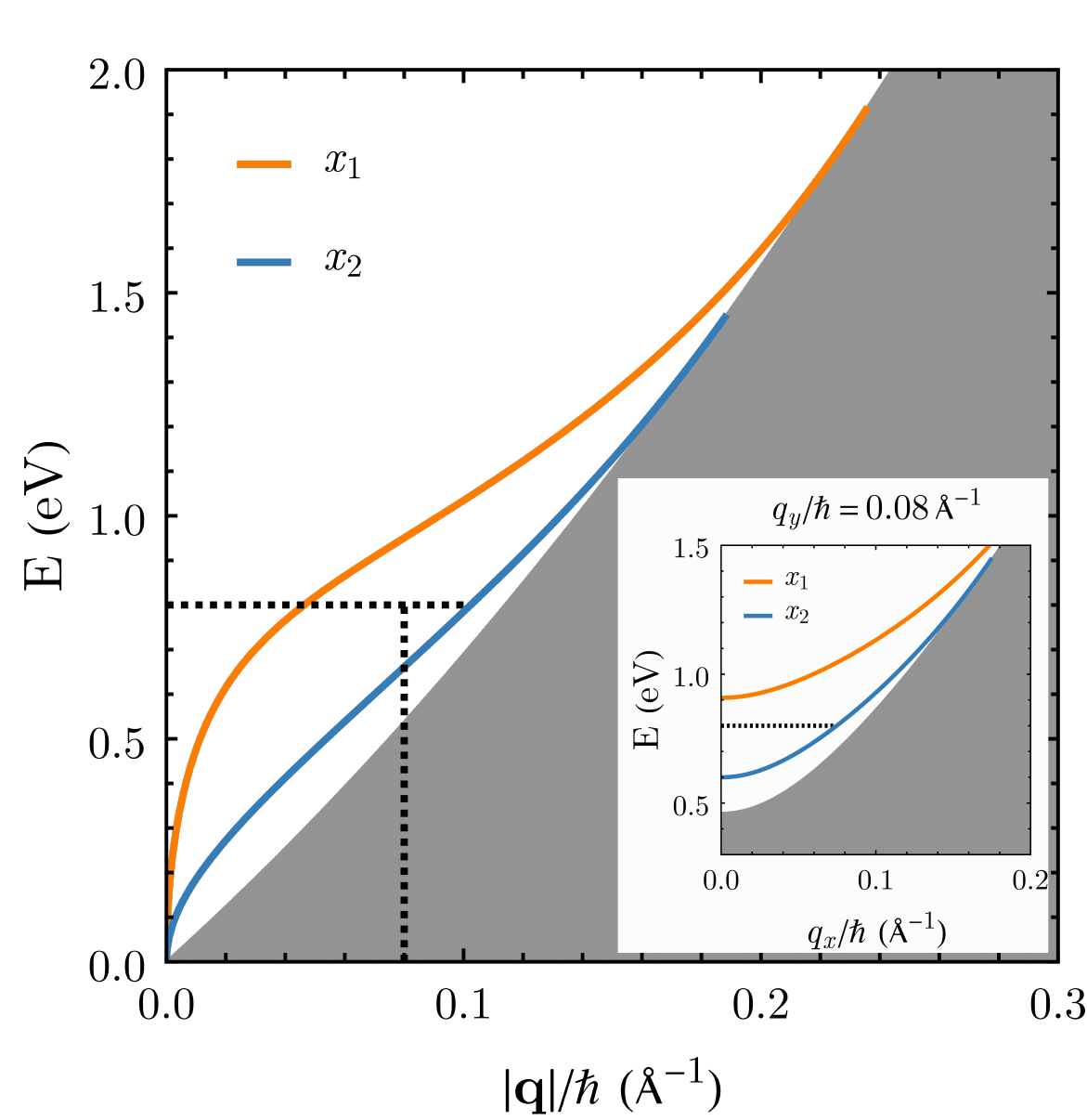}
  \caption{The local dispersion relation for plasmons in two-dimensional systems at two points with different values of the substrate dielectric constant, namely $\varepsilon_{\mathrm{b}} = 1$ at point $x_1$, and  $\varepsilon_{\mathrm{b}} = 9$ at point $x_2$. The horizontal dashed black line indicates a constant energy $E_\mathrm{pl} = 0.8$~eV, and the vertical dashed black line indicate a momentum $q_y /\hbar = 0.08$~\r{A}$^{-1}$. For this momentum, the dispersion $E(q_x)$ is plotted in the inset, where we see a gap opening up for $q_x = 0$. In the inset, the energy $E_\mathrm{pl} = 0.8$~eV lies below the dispersion for $\varepsilon_{\mathrm{b}} = 1$, and therefore in the classically forbidden region. For the region where $\varepsilon_{\mathrm{b}} = 9$, a state exists for this energy. The gray area in both plots depicts the Landau damped region or the particle hole continuum.}  \label{fig:plotDispDielectric}
\end{figure}

Figure~\ref{fig:plotDispDielectric} shows the plasmon dispersion for two different substrate dielectric constants as a function of the total momentum $|\mathbf{q}| / \hbar$. In orange, denoted by $x_1$, the plasmon dispersion is given for a system where the dielectric constant equal to $\varepsilon_\mathrm{b}=1$. In blue, denoted by $x_2$, the plasmon dispersion is given for $\varepsilon_\mathrm{b} = 9$. In general, the energies of the plasmon mode are pushed towards the electron-hole continuum (Landau damped region) for higher values of $\varepsilon_{\mathrm{b}}$, because of an increased screening by the substrate. For the following discussion, it is insightful to split the momentum $|\mathbf{q}|$ into two components, namely $|\mathbf{q}|^2 = q_x^2 + q_y^2$, where, as discussed before, $q_y$ is the momentum along the direction of the waveguide, and a constant of motion.  

In the case of photonic waveguides, it is convenient to describe different regions as ``faster" or ``slower" based on the refractive index, as the linear dispersion allows for a well-defined group velocity and consequently a unique critical angle. However, for plasmons, the highly nonlinear dispersion complicates the definition of ``fast" and ``slow" regions, making it difficult to define a unique critical angle. Instead, we adopt an energy-gap perspective. When treating $q_y$ as a parameter, we define the gap energy as $E_\mathrm{g}(x,q_y) = E(x,q_y; q_x = 0)$, which represents the lowest plasmon energy at a given $x$ for fixed $q_y$. The inset of Fig.~\ref{fig:plotDispDielectric} illustrates this concept by plotting $E(x,q_y;q_x)$ as a function of $q_x$ for finite $q_y/\hbar=0.08$~\r{A}$^{-1}$, at the two specific points discussed above. The gap energy can be extracted from this figure by looking at the limit $q_x \to 0$. 

The existence of this finite energy gap allows us to define classically allowed and forbidden regions, analogous to the 3D case, as follows. Suppose we excite a plasmon with energy $E_\mathrm{pl} = 0.8$~eV (i.e. the horizontal dashed line in Fig.~\ref{fig:plotDispDielectric}) and a (constant) finite momentum $q_y/\hbar =0.08$~\r{A}$^{-1}$ (i.e. the vertical dashed line), and we look at a point, $x_1$, where the dielectric constant is equal to $\varepsilon_\mathrm{b}=1$. We observe that plasmons are not allowed to propagate at this point $x_1$, since they satisfy $E_\mathrm{pl} = E(x_1;|\mathbf{q}|)$ only when $|\mathbf{q}|<|q_y|$, meaning that $q_x$ has to be imaginary. We therefore have exponentially damped waves, meaning that $x_1$ lies in a classically forbidden region. At the same time, plasmons can propagate at the point $x_2$, where $\varepsilon_\mathrm{b} = 9$, since they satisfy $E_\mathrm{pl}=E(x_2;|\mathbf{q}|)$ for $|\mathbf{q}|>|q_y|$, meaning that $q_x$ is real. The latter leads to the traveling waves, meaning that $x_2$ lies in a classically allowed region.

\begin{figure}[t]
  \centering
  \includegraphics[width=0.95\linewidth]{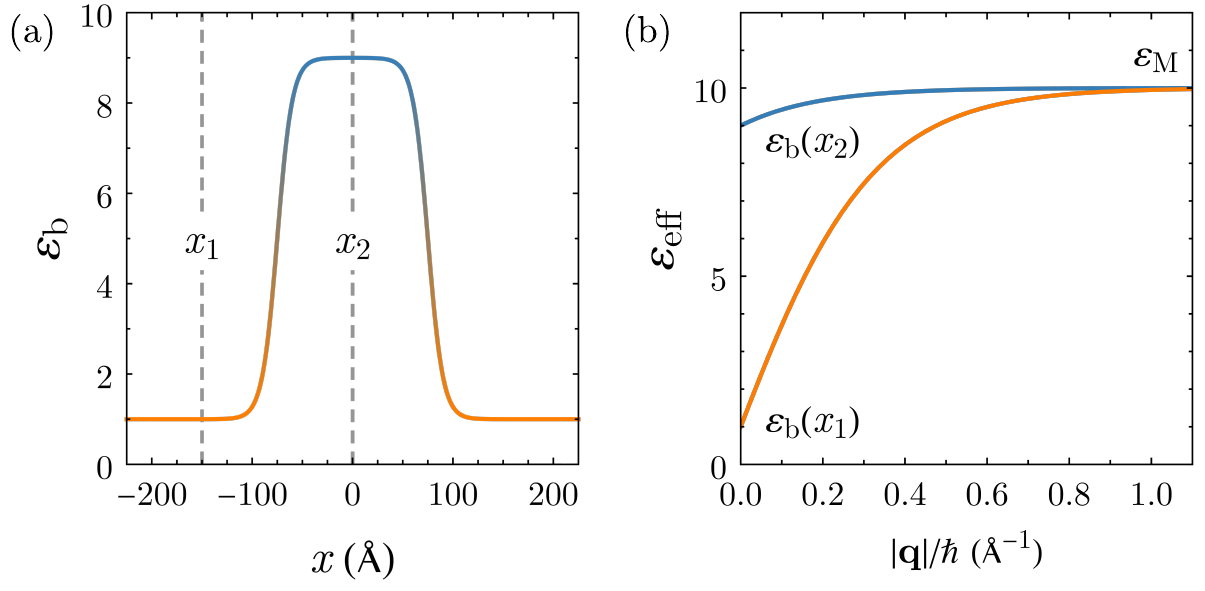}
  \caption{(a) Spatial variation of the substrate dielectric constant $\varepsilon_{\mathrm{b}} (x)$. On the left-hand and right-hand side the dielectric constant tends to $\varepsilon_{\mathrm{b}}= 1$, and in the middle it goes to $\varepsilon_{\mathrm{b}} = 9$. From Fig.~\ref{fig:plotDispDielectric}, we see that the dispersion relation in the middle region, with higher dielectric constant, is pushed towards the particle hole continuum. (b) The effective dielectric function plotted as function of $|\mathbf{q}| / \hbar$. For small $|\mathbf{q}|/ \hbar$, i.e. the long wavelength limit, the effective dielectric function goes to $\varepsilon_{\mathrm{b}}(x_i)$. In the opposite limit, for large $|\mathbf{q}| / \hbar$, it goes to $\varepsilon_{\mathrm{M}}$.
  }  \label{fig:BackgroundDielectricOverXandQ}
\end{figure}

So far, we have seen that it is possible to create  classically forbidden and allowed regions for specific energies $E_\mathrm{pl}$ and momenta $q_y$. Let us now consider the quasi-one-dimensional setup shown in Fig.~\ref{fig:BackgroundDielectricOverXandQ}(a). Ignoring the exact spatial details for the present discussion, we can clearly distinguish three different regions: on the left and the right we have $\varepsilon_\mathrm{b} = 1$ (e.g. at the point $x_1$), while $\varepsilon_\mathrm{b} = 9$ in the middle (e.g. at $x_2$). The background dielectric constant enters the classical Hamiltonian through the effective dielectric function $\varepsilon_{\mathrm{eff}}$, which is depicted in Fig.~\ref{fig:BackgroundDielectricOverXandQ}(b) as function of $|\mathbf{q}| / \hbar$ for the two values of the substrate dielectric constant $\varepsilon_{\mathrm{b}}$. For high values of $\mathbf{q}/\hbar$, the effective dielectric function tends to $\varepsilon_{\mathrm{M}}$, for both values of $\varepsilon_{\mathrm{b}}$, and the screening becomes equivalent in all regions. Comparing Figs.~\ref{fig:plotDispDielectric} and~\ref{fig:BackgroundDielectricOverXandQ}(b), we observe that, for these specific parameters, only the lower momenta in Fig.~\ref{fig:BackgroundDielectricOverXandQ}(b), are relevant for the plasmon dispersion, which means that the screening varies substantially between different regions. 

\begin{figure}[t]
  \centering
  \includegraphics[width=0.95\linewidth]{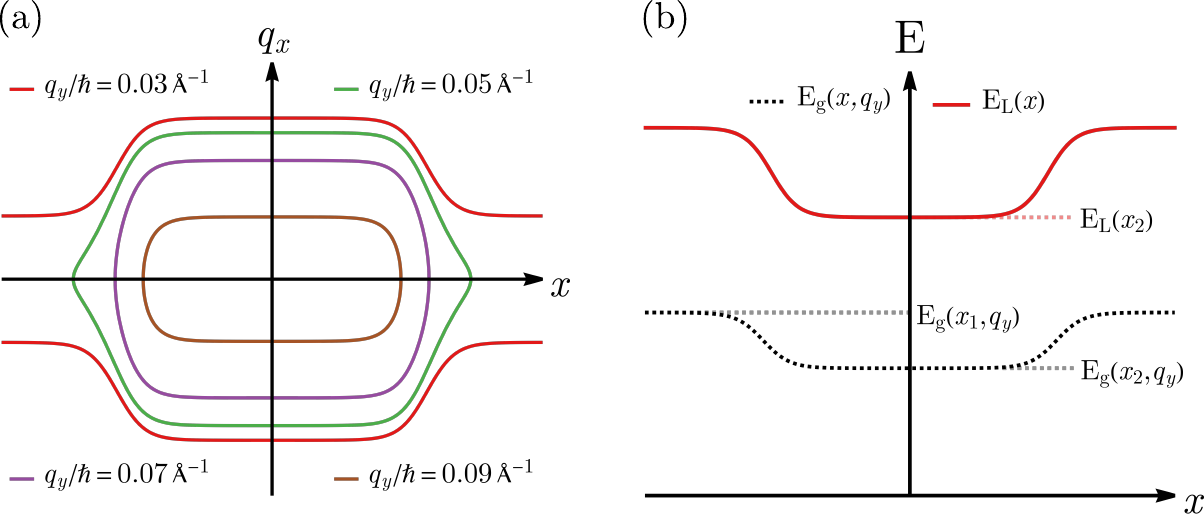}
  \caption{(a) Schematic representation of plasmonic trajectories in phase space $(x,q_x)$ for four values of the momentum $q_y$. The energy for all bound states is constant and set to $E_\mathrm{pl} = 0.8$~eV. For certain values of $q_y$, the phase space trajectories are periodic and are confined to the middle region. For the lowest value of $q_y$, we have an open trajectory, which also pervades the regions on the left and the right. (b) The spatial dependence of the gap energy $E_\mathrm{g} (x,q_y)$ (dotted black line) for constant $q_y = 0.08$~\r{A}$^{-1}$, and the energy $E_\mathrm{L}(x)$ for which the Landau damped region is reached~(solid red line). For this specific $q_y$, bound plasmonic states can exist in the middle valley of $E_\mathrm{g} (x,q_y)$. For higher values of the perpendicular momentum $q_y$, the energy $E_\mathrm{g} (x,q_y)$ increases. Above the solid red line, plasmons do not exists in either spatial region, because the dispersion has crossed the particle-hole continuum.
  }  \label{fig:plotPhaseSpaceEnergyoverX}
\end{figure}

Applying the logic of the previous paragraphs to the spatial variation $\varepsilon_\mathrm{b} (x)$ shown in Fig.~\ref{fig:BackgroundDielectricOverXandQ}(a), we conclude that, for certain energies $E_\mathrm{pl}$ and momenta $q_y$, plasmons will only be allowed to propagate in the middle region. The presence of a classically allowed region between two classically forbidden regions leads to periodic trajectories in classical phase space. These periodic trajectories are shown in Fig.~\ref{fig:plotPhaseSpaceEnergyoverX}(a), for three different values of the momentum $q_y$. Note that the size of the classically allowed region increases when $q_y$ decreases, meaning that the turning points move to larger values of $x$. Below a certain value of $q_y$, both regions are classically allowed, leading to the open trajectory that is also shown in Fig.~\ref{fig:plotPhaseSpaceEnergyoverX}(a). 

In appendix~\ref{ap:SimpleTurningPoint}, we briefly show that the turning points on the periodic trajectories, that is, the points where $q_x \to 0$, are so-called simple turning points. This means that $q_x^2 \propto x$ in the vicinity of the turning point, and holds regardless of the value of the parameter $q_y$.

Figure~\ref{fig:plotPhaseSpaceEnergyoverX}(b) can help us to gain a better understanding of the energies of which periodic trajectories occur. It shows the gap energy $E_\mathrm{g}(x,q_y)$ as function of $x$ for a specific value of $q_y/\hbar$. For this value of $q_y$, periodic trajectories can exist in the valley between $E_\mathrm{g}(x_2,q_y)<E_\mathrm{pl}<E_\mathrm{g}(x_1,q_y)$. When we increase $q_y$, we, at some point, reach a value at which the gap energy reaches the solid red line $E_{\mathrm{L}}(x)$. This corresponds to the point where the plasmon mode reaches the Landau damped region (gray area in Fig.~\ref{fig:plotDispDielectric}) Therefore, the maximum energy for which periodic trajectories can exist, is determined by the lowest Landau energy, which is for our system given by $E_{\mathrm{L},\mathrm{min}} = E_{\mathrm{L}}(x_2)$ (red dotted line in Fig.~\ref{fig:plotPhaseSpaceEnergyoverX}(b)), cf. Ref.~\cite{Reijnders22}. 

Note that, according to Ref.~\cite{Reijnders22}, so-called Landau turning points can exist for energies between $E_{\mathrm{L},\mathrm{min}} < E_\mathrm{pl} < E_{\mathrm{L},\mathrm{max}}$, where $E_{\mathrm{L},\mathrm{max}} = E_{\mathrm{L}}(x_1)$. However, these turning points and the subsequent periodic trajectories are not discussed in this paper, because this region is relatively small and close to the particle-hole continuum (Landau damped region), as can be seen in Fig.~\ref{fig:plotDispDielectric}. 

So far, we have established the existence of classically allowed and forbidden regions, and we have discussed the conditions for periodic trajectories in phase space to arise. However, not al periodic trajectories correspond to bound states. Specifically, only periodic trajectories for which the classical action fulfills the Einstein-Brillouin-Keller quantization condition~\cite{Maslov81,Guillemin77,Berry72} lead to bound states. This condition can be stated as
\begin{equation}\label{eq:quantization-condition}
  \frac{S_\mathrm{tot}}{2 \hbar} = \frac{1}{\hbar} \left|\int_{x_{c_1}}^{x_{c_2}} q_x (x) dx \right| = \left(m + \frac{1}{2}\right) \pi,
\end{equation}
where $x_{c_1}$ and $x_{c_2}$ are the classical turning points, and $m$ is a (non-negative) integer.
We can intuitively understand this condition from the requirement that the induced potential should be single-valued when we move along the periodic trajectory in phase space. After one full revolution in phase space, see Fig.~\ref{fig:plotPhaseSpaceEnergyoverX}(a), the action should have increased by a multiple of $2\pi$, which makes the potential~(\ref{eq:scansatz2d}) single-valued because it is invariant under phase differences of $2\pi$.

The factor $\pi/2$ in Eq.~(\ref{eq:quantization-condition}) accounts for the phase shift of the solution~(\ref{eq:scansatz2d}) at a simple turning point, which can be formalized through the so-called Maslov index~\cite{Maslov81,Guillemin77,Reijnders22}. This phase shift arises because the asymptotic solution~(\ref{eq:scansatz2d}) breaks down at a turning point, since the Jacobian vanishes. From a practical perspective, the Maslov index ensures the correct phase evolution as the plasmon passes through a turning point, compensating for the sign change in the Jacobian in the amplitude, see Eq.~(\ref{eq:full-potential}).
 
The quantization condition~(\ref{eq:quantization-condition}) determines the spectrum of the plasmonic waveguide. It defines a one-to-one relation between the energy and the transverse momentum $q_y$, for a given $m$. In the next subsection, we will numerically implement the waveguide discussed here and compute the spectra for waveguides with variations in different parameters.

\subsection{Numerical implementation for semiclassical bound states}\label{subsec:NumImplBoundStates}
In this subsection, we numerically demonstrate the formation of plasmonic bound states in quasi-one-dimensional systems with spatially varying parameters. We investigate three distinct scenarios: variations in the substrate dielectric constant $\varepsilon_\mathrm{b}(x)$, variations in both the dielectric constant and the electron density $n^{(0)}(x)$, and variations in the dielectric constant, electron density, and the effective height of the thin film $d(x)$. These scenarios explore how different physical mechanisms can be utilized to engineer plasmonic waveguides, i.e. invasively or non-invasively. We show that each additional degree of freedom allows for more precise control over the bound state spectrum.

For all three scenarios, we consider a quasi-one-dimensional geometry with three distinct regions: two outer regions with identical characteristics (denoted by subscript 1) and a central region with different properties (denoted by subscript 2). The spatial variation of the parameters along the $x$-direction is modeled using hyperbolic tangent functions ($\propto\pm\tanh\left[x/\ell\pm\ell_\mathrm{w}/(2 \ell)\right]$), ensuring a smooth transition between the regions. A specific example with varying the dielectric constant is given in Fig.~\ref{fig:BackgroundDielectricOverXandQ}(a), where significant variations happen over length scales $2 \ell = 3$~nm, and $\ell_\mathrm{w} = 15$~nm is the width of the middle part. The system is translationally invariant in the $y$-direction, which defines the propagation direction of the waveguide.

Throughout this section, we adopt parameters resembling a metallic system with a parabolic electronic dispersion and an effective electron mass $m_\mathrm{eff} = 0.423 m_e$. The 2D material has a background electron density $n^{(0)} = 1.8 \times 10^{14}$~cm$^{-2}$, surrounded by a thin film with an effective height $d_0 = 0.576$~nm and a dielectric constant $\varepsilon_{\mathrm{M}} = 10$, consistent with values reported in Ref.~\cite{Jiang21}. The width of the central region is set to $\ell_\mathrm{w} = 15$ nm, and the characteristic length of the boundary between regions is $2\ell = 3$ nm. These parameters yield a small dimensionless  parameter $h = \hbar / (2 \ell p_\mathrm{F}) = 0.1$, satisfying the criteria for the approximation as discussed in Refs.~\cite{Reijnders22, Koskamp23}.

\subsubsection{Varying dielectric constant}\label{subsubsec:BS-DielectricConstant}
We first consider a system with spatial variations in the substrate dielectric constant $\varepsilon_{\mathrm{b}}(x)$. The dielectric constant is varied between $\varepsilon_{\mathrm{b}}(x_1) = 1$ in the outer regions and $\varepsilon_{\mathrm{b}}(x_2) = 9$ in the central region, as described by
\begin{align}\label{eq:spatialdependenceDielectric}
  \varepsilon_{\mathrm{b}} (x) &=  \varepsilon_{\mathrm{b}}(x_1) -  \frac{\varepsilon_{\mathrm{b}}(x_1) -\varepsilon_{\mathrm{b}}(x_2)}{2} \tanh\left[\frac{x}{\ell}+ \frac{\ell_\mathrm{w}}{2 \ell}\right] \\
  &\qquad + \frac{\varepsilon_{\mathrm{b}}(x_1)-\varepsilon_{\mathrm{b}}(x_2)}{2} \tanh\left[\frac{x}{\ell} - \frac{\ell_\mathrm{w}}{2 \ell}\right], \nonumber
\end{align}
and shown in Fig.~\ref{fig:BackgroundDielectricOverXandQ}(a). This variation can be achieved non-invasively by patterning the substrate. While the hyperbolic tangent function used to model $\varepsilon_\mathrm{b}(x)$ formally only reaches its maximum value at infinity, the spatial separation of the points is sufficient for the dielectric constant to effectively reach its constant asymptotic value, ensuring that $\varepsilon_{\mathrm{b}} (x)$ is locally constant.

\begin{figure}[t]
  \centering
  \includegraphics[width=0.95\linewidth]{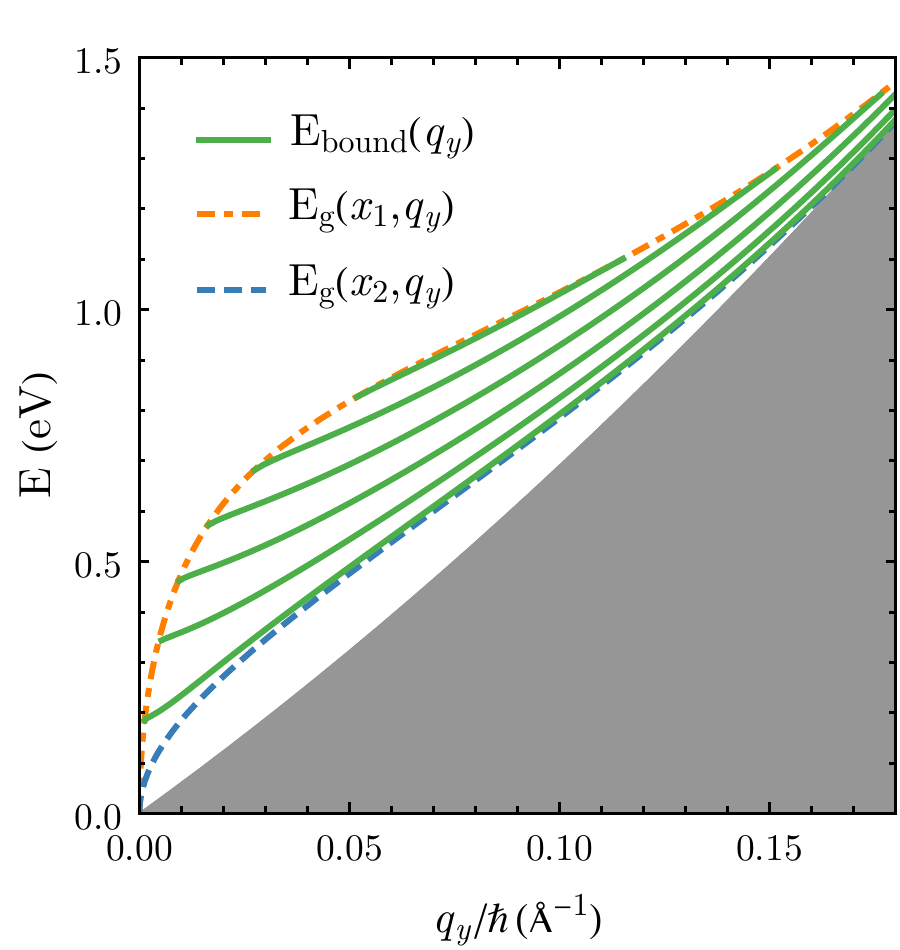}
  \caption{Spectrum of plasmonic bound states in a two-dimensional waveguide, with a spatially varying dielectric constant of the substrate, as function of the perpendicular momentum $q_y$. The dot-dashed orange and dashed blue lines indicate $E_\mathrm{g}(x_1,q_y)$ and $E_\mathrm{g}(x_2,q_y)$, respectively: they represent the upper and lower energy boundaries of the bound state spectrum. Above the energy $E_\mathrm{g}(x_1,q_y)$, the plasmons are classically allowed in both spatial regions, but, therefore, no turning points can be formed, and hence no bound states exist. }  \label{fig:BoundStatesDisp}
\end{figure}

The quantization condition Eq.~(\ref{eq:quantization-condition}) defines a unique relation between the bound state energy $E_\mathrm{bound}$ and the momentum $q_y$ for a given quantum number $m$. In Fig.~\ref{fig:BoundStatesDisp}, we show this bound state spectrum for the dielectric substrate (\ref{eq:spatialdependenceDielectric}). The green lines represent the allowed plasmon energies, with the lowest line corresponding to $m=0$. The spectrum is bounded by the gap energies $E_\mathrm{g}(x_1, q_y)$ and $E_\mathrm{g}(x_2, q_y)$, indicated by the dashed orange and blue lines, respectively. Plasmons are classically forbidden in all regions below the dashed blue curve, whilst above the dashed orange curve plasmons are classically allowed in both regions. As a consequence of the latter, plasmons above $E_\mathrm{g}(x_1,q_y)$ do not have simple turning points and are therefore in a continuum of allowed states. Bound states close to this continuum of states or to the electron-hole continuum (in gray) will probably not be measurable as localized states at finite temperatures, due to broadening of the modes into the respective continuum~\cite{Giuliani05,Kittel05}. 

This demonstrates the formation of plasmonic bound states by solely (non-invasively) manipulating the dielectric environment, which has a large effect on the plasmon dispersion. Plasmons are localized in regions of higher substrate dielectric constant. This may seem counterintuitive, as the potential (Eq.~(\ref{eq:full-potential})) naively suggests a decrease in amplitude due to increased screening. However, this screening effect does not create the classically allowed and forbidden regions necessary for bound states. We will further explore the effect of screening on the amplitude in Sec.~\ref{sec:Energy-Localized-states}.

\subsubsection{Varying dielectric constant and electron density}\label{subsubsec:BS-ElectronDensity}
We now investigate the combined effect of spatially varying both the dielectric constant $\varepsilon_\mathrm{b}(x)$ and the electron density $n^{(0)}(x)$.  Besides the dielectric variation described in the previous subsection, we introduce a $15\%$ increase in $n^{(0)}$ in the central region; while this specific value is illustrative, the qualitative effects of increased carrier density are the focus of this investigation. This increase can, for example, be achieved non-invasively through local gating or invasively through doping.

The electron density is parameterized via the Thomas-Fermi approximation, $p_{\mathrm{F}} (\mathbf{x}) = \hbar \sqrt{4 \pi n^{(0)}(\mathbf{x}) / g_\mathrm{s}}$, and its spatial variation follows the same hyperbolic tangent profile as the dielectric constant (Eq.~(\ref{eq:spatialdependenceDielectric})), where we parameterize the electron density in the middle part with a relative increase $\delta n$ compared to the background electron density, $n_0^{(0)}$ in the outer regions.

While increasing the dielectric constant tends to lower the plasmon dispersion, increasing the electron density has the opposite effect~\cite{Koskamp23}. These two competing effects operate over different momentum ranges. Consequently, we expect a crossover regime where the influence of increased electron density outweighs the increased screening from the dielectric constant.

\begin{figure}[t]
  \centering
  \includegraphics[width=0.95\linewidth]{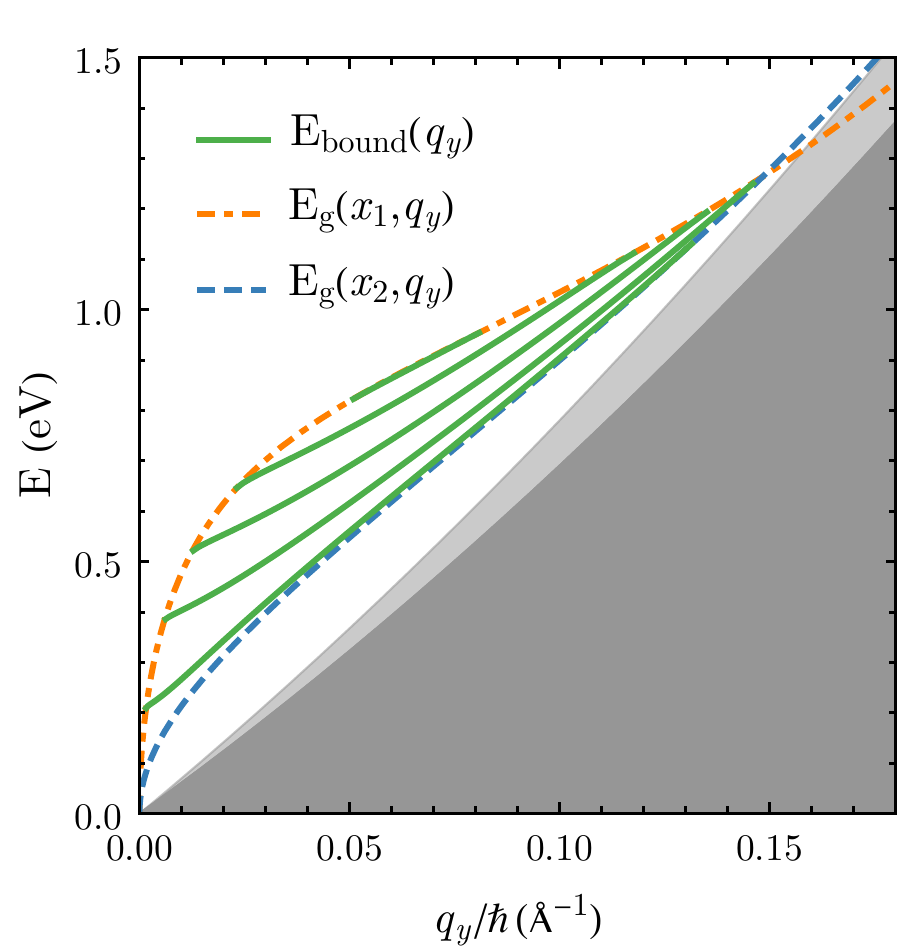}
  \caption{Spectrum of plasmonic bound states in a two-dimensional waveguide, with a spatially varying dielectric constant and electron density, as function of the perpendicular momentum $q_y$. The dot-dashed orange and dashed blue lines indicate $E_\mathrm{g}(x_1,q_y)$ and $E_\mathrm{g}(x_2,q_y)$, respectively: they represent the upper and lower energy boundaries of the bound state spectrum. Around $q_y / \hbar = 0.15$~\r{A}$^{-1}$, the two gap energies cross, due to a combination of the increased electron density and dielectric constant in the waveguide channel. Before this crossing point, we have the bound state spectrum given by the green lines, and above the energy $E_\mathrm{g}(x_1,q_y)$, we again have a continuum of plasmon states.}  \label{fig:plotBoundStatesDispersionDensityDielectric}
\end{figure}

The resulting bound state spectrum is shown in Fig.~\ref{fig:plotBoundStatesDispersionDensityDielectric}. The interplay between $\varepsilon_\mathrm{b}(x)$ and $n^{(0)}(x)$ leads to a crossing of the gap energies $E_\mathrm{g}(x_1, q_y)$ and $E_\mathrm{g}(x_2, q_y)$ at higher $q_y$.  This crossing point defines the boundary of the allowed energy and momentum ranges for the bound states and can be tuned by adjusting the magnitude of the variations in both parameters. This demonstrates the enhanced control over the bound state spectrum achieved by incorporating electron density variations.

All bound states, except for the lowest state ($m=0$), begin and end at the continuum edge defined by $E_\mathrm{g}(x_1, q_y)$ (dashed orange line). The $m=0$ state ends at $E_\mathrm{g}(x_2, q_y)$ (dashed blue line), seemingly in a classically forbidden region. However, due to the opposing influences of the varying dielectric constant and electron density, $E_\mathrm{g}(x, q_y)$ locally dips below $E_\mathrm{g}(x_2, q_y)$ in the boundary region between the central and outer regions. This local decrease arises from the distinct momentum dependence of the two effects. While this local minimum in $E_\mathrm{g}(x, q_y)$ could theoretically support a bound state localized at the boundary, we verified, through numerical estimates, that this is not the case in our setups. We therefore do not further explore this possibility in this paper.

Increasing $n^{(0)}$ not only increases the dispersion energy $E(x;|\mathbf{q}|)$ but also raises the energy of the particle-hole continuum (Landau damped region), as shown by the lighter and darker gray areas in Fig.~\ref{fig:plotBoundStatesDispersionDensityDielectric}, corresponding to the Landau damped regions at $x_2$ and $x_1$, respectively. One could theoretically consider reversing the parameter variations, placing the higher dielectric constant and electron density regions on the outside. This might lead to a lower bound on the allowed energies and momenta. However, the Landau damped region must be carefully considered in such a scenario, as for our paremeters, the energy $E_\mathrm{g}(x_1, q_y)$ crosses into the particle-hole continuum of the new outer regions, resulting in damping for those $q_y$ values.

In summary, the interplay of the varying dielectric constant and electron density creates a crossing point in the gap energies, providing control over the allowed energy and momentum ranges for the bound states. This highlights the increased flexibility in engineering the bound state spectrum by incorporating electron density variations. Both variations in the dielectric constant and the electron density can be done non-invasively. On the contrary, when the electron layer itself is varied, the characteristic properties and therefore the parameters change, e.g. the dielectric constant $\varepsilon_{\mathrm{M}}$ or the parameter $d$ for the effective height of the thin film~\cite{Steinke20,Jiang21}. 

\subsubsection{Varying dielectric constant, electron density, and the effective height $d$}\label{subsubsec:BS-EffectiveThickness}
Here, we briefly discuss the effect of varying the effective height $d(x)$, which influences the plasmon dispersion through $\exp{(-|\mathbf{q}| d /\hbar)}$ in $\varepsilon_\mathrm{eff}$, Eq.~(\ref{eq:definition-effective-dielectricconstant}). However, while variations in the dielectric constant and electron density offer significant control over the bound state spectrum, the impact of varying the effective height is generally smaller. As documented in Refs.~\cite{Steinke20,Jiang21}, typical variations in $d$ are on the order of $10\%$, resulting in negligible changes to the bound state energies.  Therefore, to illustrate the qualitative effects of varying $d$, we will consider significantly larger, and potentially unrealistic, variations.

While not physically realistic at present, these large variations serve to illustrate the sensitivity of the bound state spectrum to changes in the effective height. Such variations could become relevant in systems with significant material or structural changes in the thin film, or potentially through substrate modifications (e.g., doping) that influence the out-of-plane penetration of the electron wavefunction.

\begin{figure}[t]
  \centering
  \includegraphics[width=0.95\linewidth]{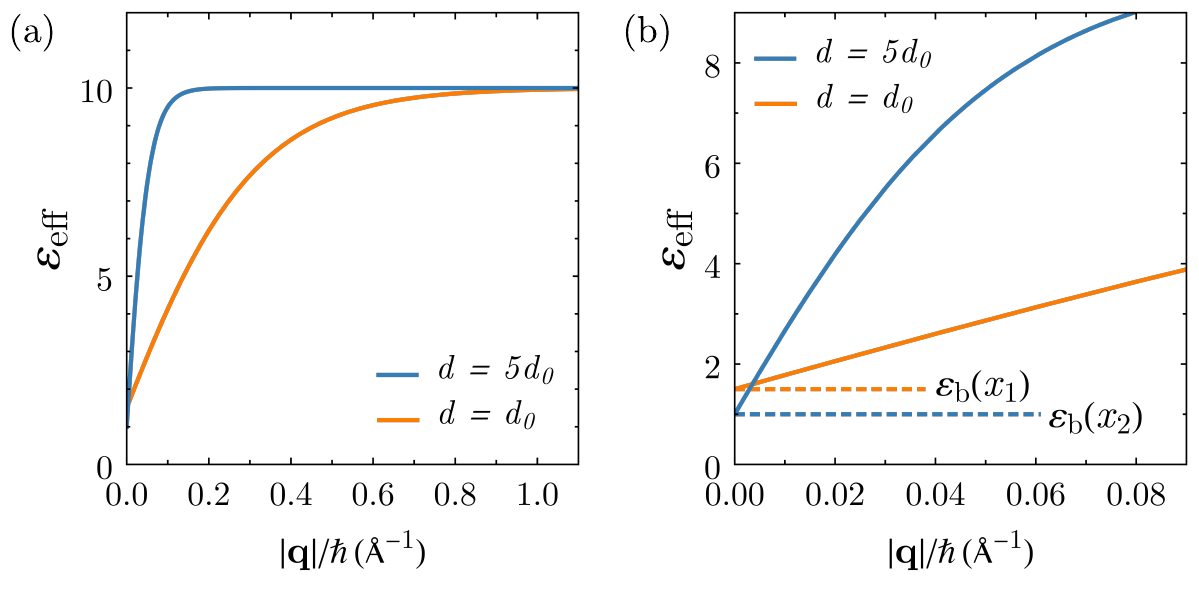}
  \caption{(a) The effective dielectric function $\varepsilon_{\mathrm{eff}}$ plotted as function of the momentum $|\mathbf{q}| / \hbar$. In the middle of the waveguide the effective thickness is equal to $d = 5 d_0$ (depicted in blue), whereas on the outer regions, the effective thickness is given by $d = d_0$ (depicted in orange) . For small $|\mathbf{q}| / \hbar$, or in the long wavelength limit, the effective dielectric function goes to $\varepsilon_{\mathrm{b}} (x_1) = 1.5$ and $\varepsilon_{\mathrm{b}} (x_2) = 1$ . In the opposite limit, for large $|\mathbf{q}| / \hbar$, it goes to $\varepsilon_{\mathrm{M}}$. The two lines cross for small $|\mathbf{q}|/\hbar$, which can be seen clearly in (b).
  }  \label{fig:dielectricoverXVaryingd}
\end{figure}

Varying the effective height alters the effective dielectric function $\varepsilon_{\mathrm{eff}}$ (Eq.~(\ref{eq:definition-effective-dielectricconstant})), influencing the screening of the plasmon. Increasing $d$ causes $\varepsilon_{\mathrm{eff}}$ to approach $\varepsilon_{\mathrm{M}}$ more rapidly as a function of momentum $|\mathbf{q}|$, as shown in Fig.~\ref{fig:dielectricoverXVaryingd}(a).  This effect is, therefore, most pronounced at lower momenta.  Furthermore, the impact of varying $d$ is greater when the difference between $\varepsilon_{\mathrm{b}}$ and $\varepsilon_{\mathrm{M}}$ is larger. Consequently, varying $d$ in regions where $\varepsilon_{\mathrm{b}}$ is already close to  $\varepsilon_{\mathrm{M}}$ (e.g., $\varepsilon_{\mathrm{b}}=9$ as in previous subsections) has a limited effect on the dispersion relation.

As mentioned above, realistic variations in $d$ (e.g., $10\%$ as discussed in Ref.~\cite{Steinke20}) result in minimal changes to the bound state spectrum.  For instance, a $10\%$ increase in $d$ leads to less than a $2.5\%$ change in energy eigenvalues where $\varepsilon_{\mathrm{b}}=1$ and only $0.1\%$ where $\varepsilon_{\mathrm{b}}=9$.

\begin{figure}[t]
  \centering
  \includegraphics[width=0.95\linewidth]{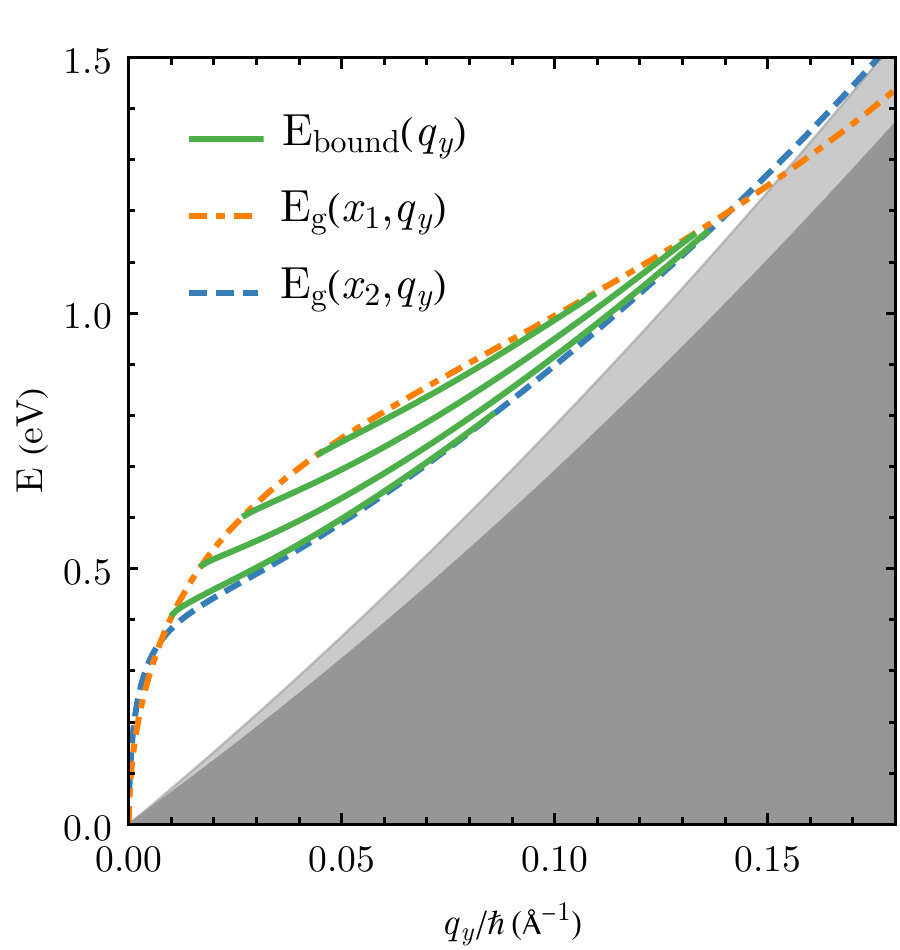}
  \caption{Spectrum of plasmonic bound states in a two-dimensional waveguide,with a spatially varying dielectric constant, electron density and effective height, as function of the perpendicular momentum $q_y$. The dot-dashed orange and dashed blue lines indicate $E_\mathrm{g}(x_1,q_y)$ and $E_\mathrm{g}(x_2,q_y)$, respectively: they represent the upper and lower energy boundaries of the bound state spectrum. The two gap energies have two crossing points, due to the interplay of the three varying parameters. The first crossing is at low momentum around $q_y / \hbar = 0.01$~\r{A}$^{-1}$, whereas the second crossing lays at higher momentum around $q_y / \hbar = 0.14$~\r{A}$^{-1}$. In between the crossing points, we have the bound states given by the green lines, and above the energy $E_\mathrm{g}(x_1,q_y)$, we again have a continuum of plasmon states.}  \label{fig:plotBoundStatesDielectricDensityThickness}
\end{figure}

However, to illustrate the potential impact of larger variations in $d$, we consider an unrealistic scenario where $d$ is increased by a factor of 5 in the central region. The effective height is parametrized in the same way as the Fermi momentum, namely with three regions where the middle layer has a relative change in height $\delta d$ and the boundary is described by a hyperbolic tangent. We set $\varepsilon_\mathrm{b} = 1$ in the central region and $\varepsilon_\mathrm{b} = 1.5$ in the outer regions. This large increase in $d$ significantly alters $\varepsilon_{\mathrm{eff}}$ at low momenta, depicted in Fig.~\ref{fig:dielectricoverXVaryingd}(b). For these low momenta, the outer regions experience stronger screening, while the central region is more screened at higher momenta. This leads to a crossing point in the gap energies $E_\mathrm{g}(x_i,q_y)$ of the two regions at low momenta, as shown in Fig.~\ref{fig:plotBoundStatesDielectricDensityThickness}.

Besides these variations, we increase the electron density by $15\%$ in the central region, as in the previous subsection. This further modifies the dispersion relation, leading to another crossing point in the gap energies at higher momenta, as can be seen in Fig.~\ref{fig:plotBoundStatesDielectricDensityThickness}. These two crossing points define the upper and lower limits for the existence of bound states, which are again depicted by green in the figure.

In summary, while realistic variations in the effective height $d$ have a negligible impact on the bound state spectrum, we have shown that unrealistically large variations can lead to significant changes, creating additional crossing points in the gap energies and thus further modifying the allowed energy and momentum ranges for bound states. This illustrates the potential, albeit in an unrealistic regime, for controlling the bound state spectrum through variations in the thin film's effective height.

\section{(Quasi)localization of plasmons through local screening}\label{sec:Energy-Localized-states}
In the previous section, we demonstrated that the effective classical Hamiltonian supports classically allowed and forbidden regions, leading to the formation of semiclassical bound states. In these states, plasmons are strictly confined within the waveguide channel and cannot propagate outside. However, effective plasmon localization can also arise through a different mechanism: variations in the local screening environment, which alters the amplitude. As explored in Ref.~\cite{Jiang21}, plasmons can be (quasi)localized even when they are classically allowed in all regions, due to variations in dielectric screening.

In this section, we investigate screening-induced plasmon (quasi)localization within our theoretical framework. We  demonstrate that while the plasmon remains classically allowed to propagate throughout the structure, variations in the dielectric environment can significantly modulate the plasmon amplitude, leading to a form of quasi-localization. Specifically, we show how the interplay of competing screening effects can be exploited to engineer plasmonic waveguides.

First, we discuss the specific setup considered throughout the section, and we present a detailed analysis of the plasmon amplitude, highlighting the different contributions to screening and their influence on the plasmon excitation. Next, we compare our results with the numerical calculations presented in Ref.~\cite{Jiang21}, demonstrating the connection between our theoretical framework and previous work. Finally, we explore potential waveguide applications, showcasing how variations in the dielectric environment can be used to control plasmon propagation through competing screening effects on the plasmon amplitude.

\subsection{The effect of screening on the amplitude}\label{subsec:Amplitude-localized-states}
In order to use the analytical theory developed in this paper, we examine the amplitude of the induced potential Eq.~(\ref{eq:full-potential}), and how it is influenced by inhomogeneities. The amplitude Eq.~(\ref{eq:Amplitude}) has three contributing factors (not considering the Berry phase) which can depend on the position, namely: the total momentum, the effective dielectric function, and the Jacobian. We will analyze the influence of these factors separately in this subsection. However, to make an accurate comparison of our results with the numerical findings in Ref.~\cite{Jiang21}, we will first define a specific setup.

Throughout this section, we consider variations in the substrate dielectric constant only.  Specifically, we analyze a system with hard-wall boundary conditions at $x_0$ and $x_\mathrm{w}$, with a total width of $240$~\r{A}. The hard-wall boundary conditions force the induced potential vanish at the boundary, leading to an Einstein-Brillouin-Keller quantization condition for the action. 

Before we consider the quantization condition, it is more insightful to construct the induced semiclassical potential, as was done in Ref.~\cite{Reijnders22,Koskamp23}. Considering plasmons with energy $E_\mathrm{pl}$ and transversal momentum $q_y$, and setting the reference point of the action to the left wall at $x_0$, we have contributions from we have contributions from both right- and left-moving plasmons. The left-moving component arises from scattering at the right wall. The full induced potential can then be written as
\begin{align}\label{eq:wavefunction-waveguide}
  V_\mathrm{pl} &= \varphi_0 (x) e^{\frac{i}{\hbar} (y-y_0) q_y} \\
  & \qquad \times  \left(e^{-\frac{i}{\hbar} \int_{x_0}^{x} q_x (x')dx'} + e^{ \frac{i}{\hbar} \int_{x_0}^{x} q_x (x')dx' - i \pi - i \Phi_{\mathrm{tot}}}\right), \nonumber
\end{align}
where the action is decomposed in Cartesian components and integrated along the plasmon trajectories. We set the arbitrary reference point $y_0 = 0$.  The phase $-i\pi$ accounts for the reflection at the hard wall, and $\Phi_\mathrm{tot} = S_\mathrm{tot}/\hbar$ represents the accumulated phase after one full revolution in phase space. Note that the right moving exponent is defined with a negative sign, because of the opposite direction of the momentum and velocity, as discussed in Ref.~\cite{Koskamp23}.

If we considered a system with hard-wall boundary conditions in both $x$ and $y$, as in Ref.~\cite{Jiang21}, then we would, from a semiclassical point of view, be considering a integrable square billiard~\cite{Stockmann07}. This leads to either ergodic (with incommensurate wavevector components) or periodic
trajectories (commensurate components). The latter can lead to bound states, for which both $q_x$ and $q_y$ are quantized. Instead, we only impose hard-wall boundary conditions in the $x$-direction. This means that we consider a physically accurate model for a waveguide, in which the momentum along the propagation direction is not quantized.

The quantization condition~\cite{Maslov81,Guillemin77,Berry72} imposed by the hard-wall is given by 
\begin{equation}\label{eq:quantization-condition-hardwall}
  \frac{S_\mathrm{tot}}{2 \hbar} = \frac{1}{\hbar} \left|\int_{x_{0}}^{x_{\mathrm{w}}} q_x (x) dx \right| = \left(m + 1\right) \pi,
\end{equation}
where the Maslov index contributes an additional phase $\pi$. In contrast to simple turning points, where the classical motion is smoothly reversed, since $q_x$ smoothly goes to zero, a hard-wall boundary imposes an abrupt momentum reversal, leading to a discontinuous phase shift and a different Maslov index. The number $m$ can again be any non-negative integer, and represents the number of nodes in the plasmonic bound state along the $x$-direction. Alternatively, we could have derived this quantization condition from the induced potential~(\ref{eq:wavefunction-waveguide}), by requiring that it vanishes at $x_0$. This shows that $\Phi_\mathrm{tot}$ is a multiple of $2\pi$, which is equivalent to Eq.~(\ref{eq:quantization-condition-hardwall}). Since we assume the plasmon to be classically allowed in all regions, the values of $m$ are bounded from above by the lowest Landau energy $E_{\mathrm{L},\mathrm{min}}$.

Note that multiple reflections from excitations far below $y_0$ should, in principle, be included. However, these reflections do not introduce new phase contributions due to the quantization condition; they only lead to a multiplicative factor in the amplitude, which contributes to the normalization.

When the plasmon is classically allowed in all regions, $q_x$ and $q_y$ are strictly real, and the exponents in Eq.~(\ref{eq:wavefunction-waveguide}) represent plane waves. The amplitude of these plane waves is determined by $\varphi_0(x)$ Eq.~(\ref{eq:Amplitude}), which captures the screening effects by the dielectric. Spatial variations in the substrate lead to a local change in screening of the electrons and therefore alter the amplitude of the excited plasmon in various ways (e.g. the inverse dependence on the effective dielectric function in Eq.~(\ref{eq:Amplitude})).

Besides the direct dependence on the effective dielectric function, the Jacobian and the total momentum are also altered by the change in dielectric environment. For the total momentum, this is evident from the dispersion relation in either region in Fig.~\ref{fig:plotDispDielectric}, where we see that for constant energy, the momentum increases for higher screening. On the other hand, the dependence of the Jacobian is more complex. It is given by
\begin{equation}
  J = 
  \det\begin{pmatrix}
    \frac{\partial x}{\partial \tau} & \frac{\partial x}{\partial \alpha}\\
    \frac{\partial y}{\partial \tau} & \frac{\partial y}{\partial \alpha}
  \end{pmatrix},
\end{equation}
where ($\tau$, $\alpha$) are parameters representing, respectively, the time evolution and initial conditions that determine the phase space trajectories governed by the Hamiltonian system defined by Eq.~(\ref{eq:effective-classical-Hamiltonian}) (see Ref.~\cite{Koskamp23} for a full discussion for a circular symmetric problem). Following the discussion in Refs.~\cite{Koskamp23,Reijnders22}, we parameterize the trajectories by $\tau$ and their $y$-coordinate at the point $x_0$. One can show~\cite{Koskamp23,Reijnders22} that this results in $\partial x/\partial \alpha = 0$ and $\partial y / \partial \alpha = 1$. These relations not only hold at the initial point where $\tau=0$, but at all points, as follows by analyzing the variational system~\cite{Koskamp23,Reijnders22}.

Thus, $J = |\partial x /\partial \tau|$, which is the group velocity in the $x$-direction. Using Hamilton's equations, the Jacobian can be written as
\begin{equation}
  J = \left|\frac{\partial \mathcal{H}_0}{\partial q_x}\left(x,\frac{\partial S}{\partial x},E\right)\right|,
\end{equation}
evaluated along the classical trajectories. This shows that the dependence of the Jacobian on local screening is complex, as it emerges through the derivatie of the effective classical Hamiltonian with respect to $q_x$. Although this dependence is difficult to analyze analytically, it can be calculated numerically.

So far, we have discussed the setup which will be considered in the following sections, and the various components of the amplitude and their influence on the plasmon excitation in different environments.  The amplitude is related to both the electrostatic energy density (Eq.~(\ref{eq:inplaneenergydensity})) and the leading order of the induced electron density (Eq.~(\ref{eq:2delecdens})), given by $n_0 (x) = \Pi_0(x) V_{\mathrm{pl}} (x)$. In the following subsections, we will consider the latter, similar to Ref.~\cite{Jiang21}, and numerically analyze systems with varying dielectric constants. For this analysis, we use the real part of $V_{\mathrm{pl}}(x)$ to calculate the induced electron density.

\subsection{The effect of local screening on the localization of the plasmonic excitation}\label{subsec:EnergyLocalizationPart1}
In the previous subsection, we discussed the various ways the plasmonic state depends on the local dielectric screening. Here, we investigate this dependence more directly by analyzing three distinct systems with varying dielectric constants on the right side, while keeping the left-side dielectric constant fixed at $\varepsilon_{\mathrm{b}} = 1$. The dispersion relation for a system with this dielectric constant is shown in Fig.~\ref{fig:plotDispDielectric}, and closely matches the dispersion in Ref.~\cite{Jiang21} (likewise for $\varepsilon_{\mathrm{b}} = 9$). The hard-wall boundary conditions quantize the allowed momenta, establishing a unique relationship between energy and momentum $q_y$ for a given mode number $m$. For consistency, we fix the energy at $E_\mathrm{pl} = 1.2$~eV and the mode number at $m = 3$ throughout this subsection, allowing $q_y$ to vary between systems.

\begin{figure}[t]
  \centering
  \includegraphics[width=\linewidth]{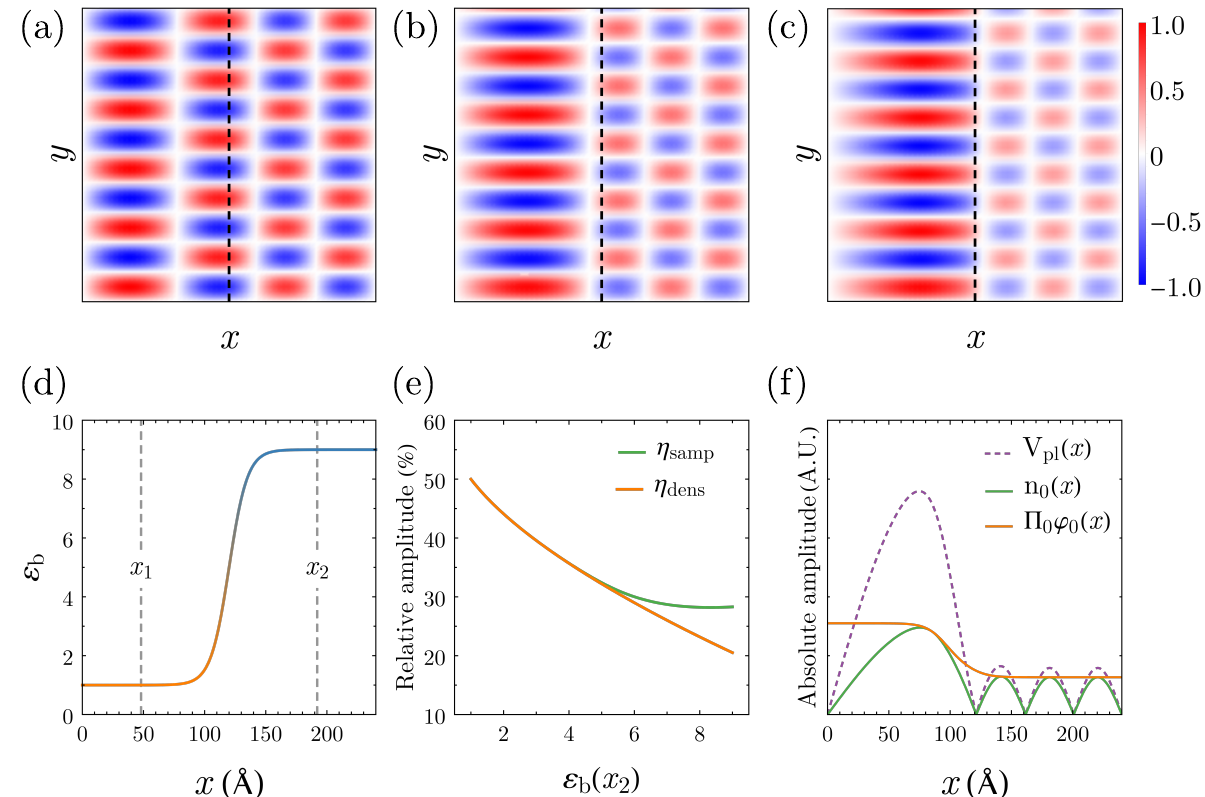}
  \caption{Localization of the plasmon in regions with lower screening. (a)-(c) Real-space induced electron density $n_0 (x,y)$, plotted for different values of the substrate dielectric constant on the right side. The left side has a dielectric constant of $\varepsilon_{\mathrm{b}}(x_1) = 1$ for all three plots, and the right side is varied, namely (a) $\varepsilon_{\mathrm{b}}(x_2) = 2$, (b) $\varepsilon_{\mathrm{b}}(x_2) = 4$, and (c) $\varepsilon_{\mathrm{b}}(x_2) = 9$. (d) Spatial variation of the dielectric constant $\varepsilon_{\mathrm{b}}(x)$, for $\varepsilon_{\mathrm{b}}(x_1) = 1$ and $\varepsilon_{\mathrm{b}}(x_2) = 9$. (e) Relative amplitude $\eta_\mathrm{dens}$ (orange) and $\eta_\mathrm{samp}$ (green) of the plasmon excitation on the right. (f) Absolute value of the induced plasmon potential, induced electron density and the amplitude of the electron density over $x$, for $\varepsilon_{\mathrm{b}}(x_2) = 6$. The orange line corresponds to the amplitude taken for $\eta_\mathrm{dens}$, and the green line for $\eta_\mathrm{samp}$.}  \label{fig:plotsDielectricLocalization}
\end{figure}

Figures~\ref{fig:plotsDielectricLocalization}(a)-\ref{fig:plotsDielectricLocalization}(c) show the real-space induced electron densities $n_0$, for the profile of the dielectric constant shown in Fig.~\ref{fig:plotsDielectricLocalization}(d). The dielectric constant on the right side takes the values $\varepsilon_{\mathrm{b}}(x_2) = (2,4,9)$, from (a) to (c).  The transition between the dielectric constants is modeled as a hyperbolic tangent, i.e. $\propto  \tanh\left[x / \ell\right] $, similar to the boundary described in Sec.~\ref{sec:bound-states}.

The electron density is clearly more localized on the left side of the system, where the dielectric constant is lower. To quantify this localization, we compare the electron density amplitude in the two regions. We define a relative amplitude as
\begin{equation}\label{eq:relativeAmplitudeDens}
  \eta_\mathrm{dens} = \frac{\Pi_0 (x_2) \varphi_0 (x_2)}{\Pi_0 (x_1) \varphi_0 (x_1) + \Pi_0 (x_2) \varphi_0 (x_2)},
\end{equation}
which is plotted in orange in Fig~\ref{fig:plotsDielectricLocalization}(e), as function of $\varepsilon_\mathrm{b} (x_2)$. Note that this relative amplitude does not include the wave-like nature of the plasmon, as it only considers the amplitude.

Figure~\ref{fig:plotsDielectricLocalization}(f), shows that the theoretical maximum of the induced electron density, $\Pi_0(x) \varphi_0(x)$, is not always attained due to the finite system size and boundary conditions. Here, the absolute value of the induced potential $V_\mathrm{pl}(x)$ (dashed purple) and the electron density $n_0(x)$ (solid green) are plotted for $\varepsilon_\mathrm{b}(x_2) = 6$. The solid orange line represents the theoretical maximum $\Pi_0(x) \varphi_0(x)$, which does not account for the wave-like features of $V_\mathrm{pl}(x)$. The maximum of the attained electron density is often located near the boundary between the two regions, particularly for higher values of $\varepsilon_\mathrm{b}(x_2)$, where half of the wavelength of the plasmon (in the $x$-direction) becomes comparable to or larger than the size of the left region.

On that note, we can also define a relative amplitude based on the maximum of the attained density in the sample, which is given by
\begin{equation}\label{eq:relativeAmplitudeSamp}
  \eta_\mathrm{samp} = \frac{n_{0,\mathrm{max}} (x_{\mathrm{left}}) }{n_{0,\mathrm{max}} (x_{\mathrm{right}}) + n_{0,\mathrm{max}} (x_{\mathrm{left}})},
\end{equation}
where the maximum of $n_0$ on the left side is taken over the left and the entire boundary region, while the region on the right side starts where the difference between $\varepsilon_{\mathrm{b}}(x)$ and $\varepsilon_{\mathrm{b}}(x_2)$ is less than $2\%$. This relative amplitude is plotted in green in Fig.~\ref{fig:plotsDielectricLocalization}(e), and shows a clear deviation from the theoretical relative amplitude $\eta_\mathrm{dens}$ from roughly $\varepsilon_\mathrm{b}(x_2) = 6$.

In summary, this subsection has demonstrated that the localization of the plasmonic excitation is significantly influenced by the local dielectric screening. By analyzing systems with varying dielectric constants, we observed that the electron density tends to localize more on the side with the lower dielectric constant, indicating that (quasi)localization of the plasmonic excitation is indeed possible under these conditions. This provides valuable insights into the behavior of plasmonic states in heterogeneous dielectric environments. We finish this discussion by noting that the real-space electron density plots and the dependence on the dielectric constant are in very good agreement with Figure 5(g) in Ref.~\cite{Jiang21}. Their numerical results appear to fall between our theoretical and attained relative amplitude for higher values of $\varepsilon_{\mathrm{b}}(x_2)$.

\subsection{Plasmonic waveguide from amplitude effect}\label{subsec:EnergyLocalizationPart2}
In the previous subsection, we demonstrated the possibility of localizing plasmon excitations in regions of lower dielectric screening, and confirmed that our calculations coincide with previous numerical results. Furthermore, Fig.~\ref{fig:plotsDielectricLocalization}(e) suggests that localization increases with greater contrast in the dielectric constant between regions. We now consider a waveguide setup analogous to those in Sec.~\ref{sec:bound-states}, but with localization occurring in the region with a lower dielectric constant. Specifically, we consider a central channel with $\varepsilon_{\mathrm{b}}(x_1) = 1$ and outer regions with $\varepsilon_{\mathrm{b}}(x_2) = 9$. The dielectric constant profile is given by Eq.~(\ref{eq:spatialdependenceDielectric}), where $x_1$ now refers to a point in the center and $x_2$ to a point in the outer regions. The central channel width is $\ell_\mathrm{w} = 100$~\r{A}, and the total system width is $240$~\r{A}, with hard-wall boundary conditions at the outer edges.  These boundary conditions imply that the induced potential goes to zero at these edges, leading to the quantization condition Eq.~(\ref{eq:quantization-condition-hardwall}),  which defines a one-to-one relation between the energy and momentum $q_y$ for a given $m$. In this subsection, we vary different parameters, such as the plasmon energy and the quantum number $m$, to see how we can control the quasi-localization for waveguiding.

\begin{figure}[t]
  \centering
  \includegraphics[width=\linewidth]{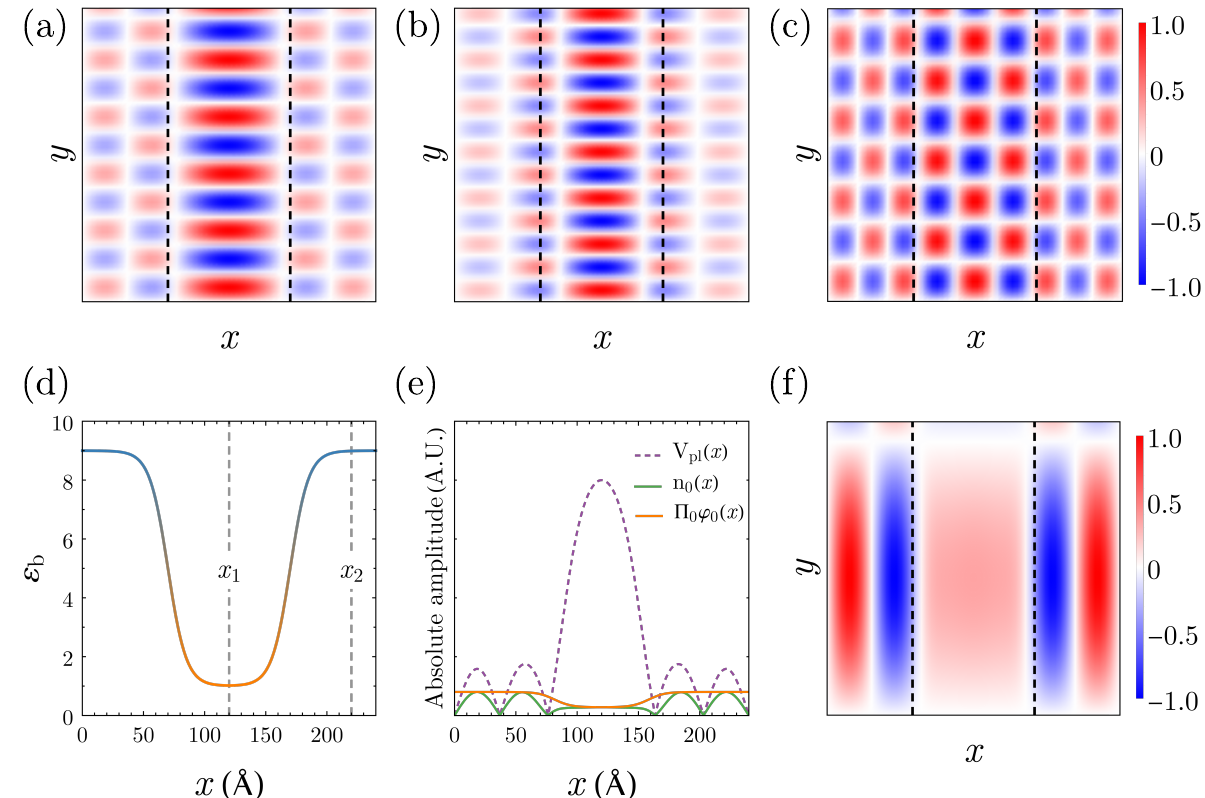}
  \caption{Localization of the plasmon in a waveguide setup. (a)-(c) Real-space induced electron density $n_0(x,y)$, plotted for different energies and number of nodes, namely (a) $E_\mathrm{pl} = 1.2$~eV and $m= 4$, (b)  $E_\mathrm{pl} = 1.4$~eV and $m= 4$, (c)  $E_\mathrm{pl} = 1.2$~eV and $m= 8$. (d) Spatial variation of the dielectric constant as a function of $x$ for the waveguide considered, where $\varepsilon_{\mathrm{b}}(x_1) = 1$ and $\varepsilon_{\mathrm{b}}(x_2) = 9$. (e) Absolute value of the induced plasmon potential $V_\mathrm{pl}$, induced electron density $n_0$ and the amplitude of the electron density $\varphi_0$ as function of $x$, for $E_\mathrm{pl} = 0.8$~eV and $m= 4$. (f) Real-space induced electron density $n_0(x,y)$, plotted for $E_\mathrm{pl} = 0.8$~eV and $m= 4$.}  \label{fig:plotsDielectricWaveguideLocalization}
\end{figure}

First, we analyze the effect of the plasmon energy. Figures~\ref{fig:plotsDielectricWaveguideLocalization}(a)-\ref{fig:plotsDielectricWaveguideLocalization}(c) show the real-space induced electron density $n_0(x,y)$ for the dielectric environment depicted in Fig.~\ref{fig:plotsDielectricWaveguideLocalization}(d). For the first two plots, (a) and (b), the number of nodes in between the hard-wall boundaries is kept constant at $m=4$, while the energy is increased from $E_\mathrm{pl} = 1.2$~eV in (a) to $1.4$~eV in (b). This energy increase results in a higher momentum $q_y$, increasing the number of nodes along the waveguide direction.We observe slightly increased localization in the central region with increasing energy.

We also consider the effect of the number of nodes $m$. In Fig.~\ref{fig:plotsDielectricWaveguideLocalization}(c), the plasmon is excited at $E_\mathrm{pl} = 1.2$~eV (as in (a)) but with a higher number of nodes in the $x$-direction, $m=8$. This leads to a lower momentum $q_y$, evident in the decreased number of nodes along the $y$-direction.  For this increase in $m$ at constant energy, the localization in the central channel decreases.

While the localization varies with energy, momentum $q_y$, and node number $m$, the electron density $n_0$ remains localized in the central channel in all three cases shown in Fig.~\ref{fig:plotsDielectricWaveguideLocalization}(a)-\ref{fig:plotsDielectricWaveguideLocalization}(c). However, at lower energies, the opposite behavior can be observed. For example, at $E_\mathrm{pl} = 0.8$~eV and $m=4$, the electron density localizes in the high-screening outer regions (Fig.~\ref{fig:plotsDielectricWaveguideLocalization}(e) and \ref{fig:plotsDielectricWaveguideLocalization}(f)). Figure~\ref{fig:plotsDielectricWaveguideLocalization}(e) shows that while the induced potential $V_\mathrm{pl}(x)$ is highly localized in the low-screening central channel, the induced electron density $n_0(x)$ is not. This suggests that this effect stems solely from the polarization $\Pi_0(x)$, which generally decreases with decreasing momentum $|\mathbf{q}|$ at constant energy. Because the momentum is lower in the central region (Fig.~\ref{fig:plotDispDielectric}), the polarization is also lower. In the higher energy cases (Fig.~\ref{fig:plotsDielectricWaveguideLocalization}(a)-\ref{fig:plotsDielectricWaveguideLocalization}(c)), the effect of the lower $\Pi_0(x)$ in the middle is less dominant; the localization is primarily driven by the increased amplitude $\varphi_0(x)$ in the low-screening region.

\begin{figure}[t]
  \centering
  \includegraphics[width=\linewidth]{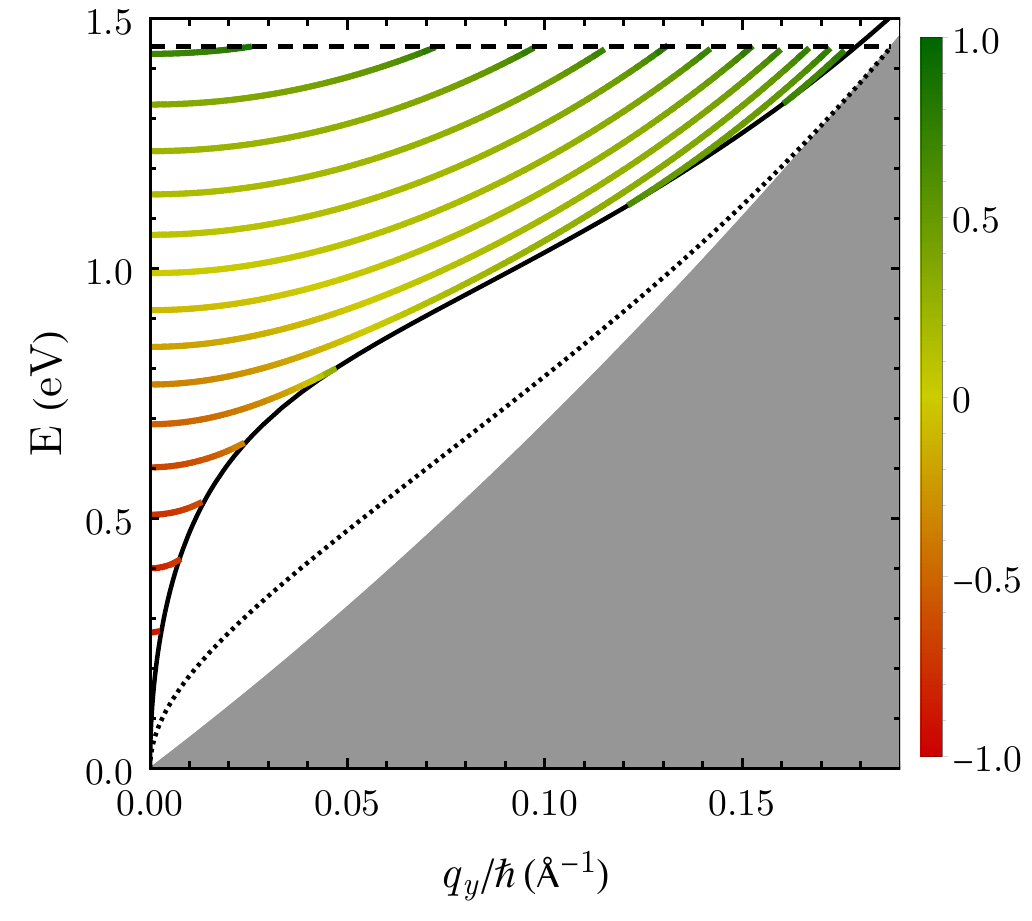}
  \caption{Spectrum of plasmonic bound states in a two-dimensional systems with a spatially varying dielectric constant as function of the perpendicular momentum $q_y$. The color gradient indicates the relative amplitude $\nu$, where localization in the low-screened middle region is depicted with green and in the high-screened edge regions with red. The solid and dotted black lines indicate $E_\mathrm{g}(x_1,q_y)$ and $E_\mathrm{g}(x_2,q_y)$, respectively. Above the energy $E_\mathrm{g}(x_1,q_y)$, in the region where we are interested in, the plasmons are allowed in both spatial regions. In this case, there is no continuum of states, due to the hard-wall boundary conditions. Above the horizontal dashed black line on top, the Landau damped region is reached in the outer regions, which starts from the energy $E_{\mathrm{L},\mathrm{min}} = E_\mathrm{L} (x_2)$. }  \label{fig:plotBoundStatesDispersionDielectricWaveguide}
\end{figure}

To get an overview of the quasi-localization and its dependence on $E$, $q_y$, and $m$, we plot the bound state spectrum in Fig.~\ref{fig:plotBoundStatesDispersionDielectricWaveguide}, similar to the plots in Sec.~\ref{sec:bound-states}, but now with states above the energy $E_\mathrm{g}(x_1,q_y$). Each curve corresponds to a different value of $m$, starting with $m=0$ at the bottom. The color gradient on each curve represents the relative amplitude difference between the central and outer regions, as given by 
\begin{equation}
  \nu = \frac{\Pi_0 (x_1) \varphi_0 (x_1) - \Pi_0 (x_2) \varphi_0 (x_2)}{\Pi_0 (x_1) \varphi_0 (x_1) + \Pi_0 (x_2) \varphi_0 (x_2)}.
\end{equation}
Positive (green) values of $\nu$ indicate localization in the central, low-screening region, while negative (red) values indicate localization in the high-screening outer regions.

As discussed earlier, the Jacobian has a complex influence on the amplitude, depending on the derivative of the effective classical Hamiltonian with respect to $q_x$.  Near the gap energy $E_\mathrm{g}(x_1,q_y)$ (solid black curve in Fig.~\ref{fig:plotBoundStatesDispersionDielectricWaveguide}), where $q_x \to 0$ in the central channel, the Jacobian is generally lower, enhancing localization in the middle. This is visible, for example, at the end of the $m=4$ bound state curve.  Furthermore, near the Landau energy $E_\mathrm{L}(x_2)$ (horizontal dashed black line), the Jacobian becomes large in the outer regions, reducing the plasmon excitation there and increasing central localization.

An analog to the bound state spectrum as function of the momentum $q_y$, given in Fig.~\ref{fig:plotBoundStatesDispersionDielectricWaveguide}, can be found in Ref.~\cite{Jiang21} figure 4(b). However, direct comparison is difficult, due to differences in system setup, with their system resembling the system discussed in the previous subsection. Besides, there is no clear distinction between the number of nodes considered. It is likely that the curve in their plot crosses multiple values of $m$ along the dots.

We have thus seen that (quasi)localization of plasmons in waveguide setups is possible (Fig.~\ref{fig:plotsDielectricWaveguideLocalization}). The outcome, however, is not straightforward and depends on many different parameters, as is evident from Fig.~\ref{fig:plotBoundStatesDispersionDielectricWaveguide}. By varying parameters, such as the dielectric constant, plasmon energy, and mode number, we can control the degree of localization. This tunability allows us to achieve either strong or weak (quasi)localization, depending on the desired application.

\section{Conclusion and outlook}\label{sec:Conclusion}
We have developed a comprehensive theory, using semiclassical techniques, to describe two-dimensional plasmonic waveguides. This theory allows us to analyze two distinct types of localized plasmonic states: semiclassical bound states and quasi-localized states arising from a varying amplitude. Our semiclassical approach enables the calculation of the full quantum plasmon dispersion within the random phase approximation, without limitations on momentum or system size. However, the theory is subject to the constraint that the characteristic length scale of variations in the inhomogeneity, $\ell$, must be much larger than the electron wavelength, $\lambda_{\mathrm{el}}$. This condition ensures the validity of the semiclassical expansion in terms of the small dimensionless parameter $h$.

Specifically, we constructed a general theory for plasmons in inhomogeneous two-dimensional systems embedded in an arbitrary three-dimensional dielectric environment. We achieved this by adiabatically separating the in-plane and out-of-plane variables and solving the resulting systems separately. The out-of-plane system yielded a differential equation, which we solved using the method of variation of parameters, leading to the effective classical Hamiltonian (Eq.~(\ref{eq:effective-classical-Hamiltonian})).  The in-plane system was treated with semiclassical techniques, thus being subject to the limit of small $h$ (for a full discussion of the semiclassical approximation and its applicability to plasmonic systems, see Refs.~\cite{Koskamp23,Reijnders22}).

A central result of our theory is the effective classical Hamiltonian (Eq.~(\ref{eq:effective-classical-Hamiltonian})), which incorporates the effective dielectric function (Eq.~(\ref{eq:definition-effective-dielectricconstant})) in terms of the determinant of the matrix $W$, which resembles the Wronskian determinant. This effective dielectric function can be momentum-dependent and captures screening effects from the three-dimensional dielectric environment. The classical trajectories of the plasmon in phase space can be derived from this Hamiltonian. Furthermore, it determines the classical action, $S(\mathbf{x})$, through the Hamilton-Jacobi equation, which governs the phase of the induced potential, $V_\mathrm{pl}$. We derived an explicit expression for the amplitude of $V_\mathrm{pl}$, and discussed its relation to the electrostatic energy density in Sec.~\ref{subapp:energydensity}.

We applied this theory to a specific, commonly used, model for the dielectric constant~\cite{Jiang21,Emelyanenko08,Wehling11}, representing a layered structure consisting of a thin film with effective height $d$, and dielectric constant $\varepsilon_{\mathrm{M}}$, encapsulated by a dielectric substrate on both sides. An infinitely thin 2D free electron layer was placed in the middle of the thin film. Within our framework, the dielectric constants, effective height, and electron density can all be treated as spatially varying parameters. This model facilitated the analysis of two distinct types of localized states.

The first type of localization, through the formation of bound states, arises from the effective classical Hamiltonian. These states occur when a classically allowed region is surrounded by classically forbidden regions. For 2D plasmons, the formation of these regions is strongly dependent on the momentum $q_y$. This localization mechanism is analogous to total internal reflection in optical waveguides. We numerically implemented this type of bound state in Sec.~\ref{sec:bound-states} for various spatially varying parameters (dielectric constant, electron density, and effective height $d$). Simultaneous variation of these parameters demonstrated control over the lower and upper limit of the bound state spectrum in terms of energies and momenta $q_y$. These spectra, plotted as a function of the momentum along the waveguide, $q_y$, which can be controlled experimentally by plasmonic antennas, provide a direct link to experimental observables. 

We note that these localized states are single states, meaning only one energy is excited per $m$ for a certain momentum $q_y$. For applications in integrated photonics, it would be interesting to describe a plasmonic band structure, where one can excite multiple states per quantum number $m$ for a certain momentum $q_y$. Such a band structure might exist in periodic systems or plasmonic crystals, analogous to electronic band structures. We anticipate such band structures to also depend on $q_y$.

The second type of (quasi)localization, studied in Sec.~\ref{sec:Energy-Localized-states}, arises from variations in dielectric screening, which modulate the amplitude of the plasmon and therefore the induced electron density. This (quasi)localization is purely an amplitude effect, distinct from the phase-related localization of bound states. The underlying mechanism is complex due to the interplay of several factors in the amplitude: the Jacobian, the effective dielectric function, and the total momentum. Furthermore, the polarization, $\Pi_0$, can lead to counter-intuitive localization behavior at low momenta, where localization shifts to regions of higher screening. In our analysis, we focused on variations in the dielectric constant. However, similar effects are expected from variations in other parameters, such as the electron density, as these also influence the dispersion and thus the amplitude. The results presented in this section are in good agreement with numerical results from Ref.~\cite{Jiang21}.

We remark that we only plotted the real-space induced potential for the (quasi)localized states in Sec.~\ref{sec:Energy-Localized-states}, and not for the bound states in Sec.~\ref{sec:bound-states}. The latter requires a different approach, since the induced potential~\ref{eq:scansatz2d} diverges at the classical turning points, due to the vanishing of the Jacobian. In other words, the asymptotic solution~\ref{eq:scansatz2d} no longer accurately describes the true solution. An accurate description in the vicinity of turning points could be obtained using an analogous construction in momentum space~\cite{Maslov81,Guillemin77,Reijnders22}. This would result in an expression for the induced potential that involves the Airy function. However, this construction has currently not yet been performed for plasmons in inhomogeneous systems and would require additional lengthy derivations~\cite{Reijnders22}. We believe that this would be an interesting future direction, both from a practical and a fundamental point of view.

In this article, we did not consider the influence of the Berry phase~(\ref{eq:Berry-phase-definition}), since it vanishes for the quadratic dispersion considered here. However, it would be interesting to investigate different Hamiltonians, i.e. different materials, in which this phase could play a significant role. This likely requires matrix Hamiltonians, which have an internal degree of freedom, cf. the discussion of the Berry phase for Dirac systems in Ref.~\cite{Reijnders18}. In particular, the Berry phase could strongly influence the different quantization conditions, as it introduces an additional phase term.

\section*{Acknowledgments}
We would like to thank Malte R\"osner for stimulating discussions. 
The work was supported by the European Union's Horizon 2020 research and innovation programme under European Research Council synergy grant 854843 ``FASTCORR".

\appendix

\begin{widetext}
  
\section{Additional derivations for the general description}
\label{ap:AppendixChi1}

\subsection{Derivation of subprincipal symbol $\Gamma_1$ at $z=0$}
\label{subapp:amplitude}

In this appendix, we compute the integral~(\ref{eq:generalsol-Gamma1}) explicitly at the point $z=0$, using integration by parts. We show that the final result can be cast in the form~(\ref{eq:AmplitudeExponent}).

Starting from Eqs.~(\ref{eq:generalsol-Gamma1}) and~(\ref{eq:def-f1}), we have
\begin{equation}\label{eq:gamma1forz0}
  \Gamma_1(z=0) = -4 \pi e^2 \Pi_1  g(0,0) + \int_{-\infty}^\infty g(0,z') f_{1s}(z')d z',
\end{equation}
where $g(z,z')$ is given by Eq.~(\ref{eq:generalsol-Green}) and $f_{1s}$ by Eq.~(\ref{eq:def-f1s}). The latter quantity is rather involved and contains $F_0$, $F_1$ and $\Gamma_0$. We first note that $F_1$ does not contain any differential operators in $z$, whereas $F_0$ does, see Eqs.~(\ref{eq:leading-ordersymbolODE}) and~(\ref{eq:subleading-ordersymbolODE}). To shorten the notation, we write
\begin{equation}
  F_0\bigg(\mathbf{x},\mathbf{q},z,\frac{\partial}{\partial z}\bigg) = \hat{F}_0,
\end{equation}
and omit the arguments $(\mathbf{x},\mathbf{q})$ throughout most of this appendix.

Given the structure of $g(0,z')$, see Eq.~(\ref{eq:generalsol-Green}), we can split the integration in Eq.~(\ref{eq:gamma1forz0}) into two parts, over the intervals $(-\infty,0]$ and $[0,\infty)$. We start with the computation of the integral over the interval $[0,\infty)$. From Eqs.~(\ref{eq:gamma1forz0}),~(\ref{eq:generalsol-Green}) and~(\ref{eq:def-f1s}), and applying the product rule to the second term in $f_{1s}$, we have
\begin{multline}
  \int_0^{\infty} g(0,z') f_{1s}(z')dz' 
    = \frac{w_2(0)}{\det(W)} \frac{2 \pi e^2 \hbar}{\varepsilon_{\mathrm{eff}}|\mathbf{q}| }\frac{\Pi_0}{w_1(0)} 
    \bigg( 
      \int_0^{\infty} w_1(z') F_1(z') w_1(z') dz' 
      - i \int_0^{\infty} w_1(z') \sum_j \frac{\partial \hat{F}_0}{\partial q_j} \frac{\partial w_1}{\partial x_j}(z') dz'
    \bigg) \\
    - i \frac{w_2(0)}{\det(W)} \sum_j \frac{\partial }{\partial x_j}\left(\frac{2 \pi e^2 \hbar}{\varepsilon_{\mathrm{eff}}|\mathbf{q}|} \frac{\Pi_0}{w_1(0)}\right) \int_0^{\infty}w_1(z') \frac{\partial \hat{F}_0}{\partial q_j} w_1(z') dz'. \label{eq:upperhalfinhomogeneous-step1}
\end{multline}
We proceed by removing $F_1$ from the above expression.
Since $w_i$, where $i=\{1,2\}$, is a solution of the homogeneous differential equation~(\ref{eq:homogeneousODE}), we have $\hat{F}_0 w_i = 0$. Taking a mixed partial derivative of this relation with respect to $x_j$ and $q_j$, we have
\begin{equation}\label{eq:relationF1F0afterproduct}
  \frac{i}{2}\bigg( \sum_j \hat{F}_0 \frac{\partial^2  w_i}{\partial x_j \partial q_j} + \sum_j \frac{\partial \hat{F}_0}{\partial x_j} \frac{\partial  w_i}{\partial q_j} + \sum_j \frac{\partial \hat{F}_0}{\partial q_j} \frac{\partial  w_i}{\partial x_j}\bigg)
    = -\frac{i}{2} \sum_j \frac{\partial^2 \hat{F}_0}{\partial x_j \partial q_j} w_i
    = F_1 w_i ,
\end{equation}
where the last equality holds by virtue of the last equality in Eq.~(\ref{eq:subleading-ordersymbolODE}).
With this last relation, we can remove $F_1$ from Eq.~(\ref{eq:upperhalfinhomogeneous-step1}), and obtain
\begin{align}
  \int_0^{\infty} g(0,z') f_{1s}(z')dz' 
  &= \frac{i}{2} \frac{w_2(0)}{\det(W)} \frac{2 \pi e^2 \hbar}{\varepsilon_{\mathrm{eff}}|\mathbf{q}| }\frac{\Pi_0}{w_1(0)} \sum_j
  \int_0^{\infty} w_1(z') 
  \bigg( \hat{F}_0 \frac{\partial^2  w_1}{\partial x_j \partial q_j}(z') + 
  \frac{\partial \hat{F}_0}{\partial x_j} \frac{\partial  w_1}{\partial q_j}(z')
  - \frac{\partial \hat{F}_0}{\partial q_j} \frac{\partial  w_1}{\partial x_j}(z') \bigg) dz'\nonumber\\
  &\qquad - i \frac{w_2(0)}{\det(W)} \sum_j \frac{\partial }{\partial x_j}\left(\frac{2 \pi e^2 \hbar}{\varepsilon_{\mathrm{eff}}|\mathbf{q}|} \frac{\Pi_0}{w_1(0)}\right) \int_0^{\infty}w_1(z') \frac{\partial \hat{F}_0}{\partial q_j} w_1(z') dz'. \label{eq:upperhalfinhomogeneous}
\end{align}
Although this expression looks more complicated than our initial expression~(\ref{eq:upperhalfinhomogeneous-step1}), it no longer contains $F_1$.

In what follows, we show how we can evaluate each of the terms in Eq.~(\ref{eq:upperhalfinhomogeneous}) using integration by parts. When we look at the first term in Eq.~(\ref{eq:upperhalfinhomogeneous}), we observe that it contains the product $w_1 \hat{F}_0 (\partial^2  w_1/\partial x_j \partial q_j)$. The idea of our procedure is to transfer the differential operator $\hat{F}_0$ directly to $w_1$, at the cost of a few boundary terms, and then to use that $\hat{F}_0 w_1 = 0$, which holds because $w_1$ is a solution of the homogeneous differential equation.
Explicitly, we have
\begin{align}
  &\int_0^{\infty} w_1 \hat{F}_0 \frac{\partial^2  w_1}{\partial x_j \partial q_j} dz' 
    = \int_0^{\infty}  w_1 \frac{\partial}{\partial z} \left( \varepsilon \frac{\partial}{\partial z}\right) \frac{\partial^2  w_1}{\partial x_j \partial q_j}dz' - \int_0^{\infty}  w_1 \frac{|\mathbf{q}|^2}{\hbar^2} \varepsilon \frac{\partial^2  w_1}{\partial x_j \partial q_j} dz' \nonumber \\
  & \qquad 
    = \left[ w_1 \varepsilon \frac{\partial}{\partial z} \left( \frac{\partial^2  w_1}{\partial x_j \partial q_j} \right) \right]_0^\infty 
    - \int_0^{\infty} \frac{\partial w_1}{\partial z} \varepsilon \frac{\partial}{\partial z} \left(\frac{\partial^2  w_1}{\partial x_j \partial q_j} \right) dz' 
    - \int_0^{\infty} \frac{|\mathbf{q}|^2}{\hbar^2} \varepsilon w_1(z') \frac{\partial^2  w_1}{\partial x_j \partial q_j} dz'\nonumber\\
  & \qquad 
    = \left[ w_1 \varepsilon \frac{\partial}{\partial z} \left( \frac{\partial^2 w_1}{\partial x_j \partial q_j} \right) \right]_0^\infty
    - \left[ \frac{\partial w_1}{\partial z} \varepsilon \frac{\partial^2  w_1}{\partial x_j \partial q_j} \right]_0^\infty 
    + \int_0^{\infty} \left( \frac{\partial}{\partial z} \left( \varepsilon \frac{\partial w_1}{\partial z} \right) - \frac{|\mathbf{q}|^2}{\hbar^2} \varepsilon w_1 \right) \frac{\partial^2 w_1}{\partial x_j \partial q_j} dz' ,
  \label{eq:integration-by-parts-ex}
\end{align}
where we omitted the argument $z'$ throughout. The last term now contains $\hat{F}_0 w_1$, see Eq.~(\ref{eq:leading-ordersymbolODE}), which vanishes because $w_1$ solves the homogeneous equation~(\ref{eq:homogeneousODE}). This means that the integral is given by the two boundary terms. We now recall that we assumed that $\varepsilon$ becomes constant for $z\to\infty$, and constructed our homogeneous solutions in such a way that $w_1$ decays exponentially as $z\to\infty$. This implies that all of its derivatives also vanish in this limit, and hence the boundary terms above vanish at infinity.
We therefore have
\begin{equation}  \label{eq:upperhalfinhomogeneous-first-term}
  \sum_j \int_0^{\infty} w_1 \hat{F}_0 \frac{\partial^2  w_1}{\partial x_j \partial q_j} dz'
    = - \sum_j w_1 \varepsilon \frac{\partial}{\partial z} \left(\frac{\partial^2 w_1}{\partial x_j \partial q_j}\right)
      + \sum_j \frac{\partial w_1}{\partial z} \varepsilon \left(\frac{\partial^2 w_1}{\partial x_j \partial q_j}\right) ,
\end{equation}
where all functions are to be evaluated at $z=0$.

Let us now consider the last term in Eq.~(\ref{eq:upperhalfinhomogeneous}). Since $\hat{F}_0 w_1 = 0$, we have $(\partial \hat{F}_0/\partial q_j) w_1 + \hat{F}_0 (\partial w_1/\partial q_j) = 0$, and
\begin{equation}  \label{eq:upperhalfinhomogeneous-last-term}
  \int_0^{\infty}w_1 \frac{\partial \hat{F}_0}{\partial q_j} w_1 dz'
    = - \int_0^{\infty}w_1 \hat{F}_0 \frac{\partial w_1}{\partial q_j} dz'
    = - \left[ w_1 \varepsilon \frac{\partial}{\partial z} \left( \frac{\partial w_1}{\partial q_j} \right) \right]_0^\infty
    + \left[ \frac{\partial w_1}{\partial z} \varepsilon \frac{\partial w_1}{\partial q_j} \right]_0^\infty ,
\end{equation}
where the last equality follows from repeated integration by parts, just as in Eq.~(\ref{eq:integration-by-parts-ex}).

We now show that the two remaining terms in Eq.~(\ref{eq:upperhalfinhomogeneous}) are equal to
\begin{multline}  \label{eq:upperhalfinhomogeneous-middle-terms}
  \int_0^{\infty} w_1 \frac{\partial \hat{F}_0}{\partial x_j} \frac{\partial  w_1}{\partial q_j}  dz'
  - \int_0^{\infty} w_1 \frac{\partial \hat{F}_0}{\partial q_j} \frac{\partial  w_1}{\partial x_j} dz' 
  = \left[ \frac{\partial}{\partial z} \left( \frac{\partial w_1}{\partial q_j} \right) \varepsilon \frac{\partial w_1}{\partial x_j} \right]_0^\infty 
    + \left[ w_1 \frac{\partial \varepsilon}{\partial x_j} \frac{\partial}{\partial z}\left( \frac{\partial w_1}{\partial q_j} \right) \right]_0^\infty
    - \left[ \frac{\partial w_1}{\partial z} \frac{\partial \varepsilon}{\partial x_j} \frac{\partial w_1}{\partial q_j} \right]_0^\infty \\
    - \left[ \frac{\partial}{\partial z} \left( \frac{\partial w_1}{\partial x_j} \right) \varepsilon \frac{\partial w_1}{\partial q_j} \right]_0^\infty .
\end{multline}
First, note that $\partial \hat{F}_0/\partial q_j = 2 \varepsilon q_j/\hbar^2$ is not a differential operator, which implies that
\begin{align}
  \int_0^{\infty} w_1 \frac{\partial \hat{F}_0}{\partial q_j} \frac{\partial  w_1}{\partial x_j} dz'
  &= \int_0^{\infty} \bigg( \frac{\partial \hat{F}_0}{\partial q_j} w_1 \bigg) \frac{\partial  w_1}{\partial x_j} dz'
   = - \int_0^{\infty} \bigg( \hat{F}_0 \frac{\partial w_1}{\partial q_j} \bigg) \frac{\partial  w_1}{\partial x_j} dz'  \nonumber \\
  &= - \left[ \varepsilon \frac{\partial}{\partial z}\bigg( \frac{\partial w_1}{\partial q_j} \bigg) \frac{\partial w_1}{\partial x_j} \right]_0^\infty
      + \int_0^\infty \varepsilon \frac{\partial}{\partial z}\left( \frac{\partial w_1}{\partial q_j} \right) \frac{\partial}{\partial z}\left( \frac{\partial w_1}{\partial x_j} \right) d z'
      + \int_0^\infty \frac{\partial w_1}{\partial q_j} \varepsilon \frac{q^2}{\hbar^2} \frac{\partial w_1}{\partial x_j} d z' ,
  \label{eq:upperhalfinhomogeneous-third-term}
\end{align}
where the second equality follows from the text above Eq.~(\ref{eq:upperhalfinhomogeneous-last-term}), and the last equality follows from integration by parts. Computing $\partial \hat{F}_0/\partial x_j$ explicitly, repeatedly integrating by parts, and using that $(\partial \hat{F}_0/\partial x_j) w_1 + \hat{F}_0 (\partial w_1/\partial x_j) = 0$, we also find that
\begin{align}
  \int_0^{\infty} w_1 \frac{\partial \hat{F}_0}{\partial x_j} \frac{\partial  w_1}{\partial q_j}  dz' 
  &= \left[ w_1 \frac{\partial \varepsilon}{\partial x_j} \frac{\partial}{\partial z}\left( \frac{\partial w_1}{\partial q_j} \right) \right]_0^\infty
  - \left[ \frac{\partial w_1}{\partial z} \frac{\partial \varepsilon}{\partial x_j} \frac{\partial w_1}{\partial q_j} \right]_0^\infty
  + \int_0^\infty \bigg( \frac{\partial \hat{F}_0}{\partial x_j} w_1 \bigg) \frac{\partial w_1}{\partial q_j} d z' 
  \nonumber \\
  &= \left[ w_1 \frac{\partial \varepsilon}{\partial x_j} \frac{\partial}{\partial z}\left( \frac{\partial w_1}{\partial q_j} \right) \right]_0^\infty
  - \left[ \frac{\partial w_1}{\partial z} \frac{\partial \varepsilon}{\partial x_j} \frac{\partial w_1}{\partial q_j} \right]_0^\infty
  - \int_0^\infty \bigg( \hat{F}_0 \frac{\partial w_1}{\partial x_j} \bigg) \frac{\partial w_1}{\partial q_j} d z'
  \nonumber \\
  &= \left[ w_1 \frac{\partial \varepsilon}{\partial x_j} \frac{\partial}{\partial z}\left( \frac{\partial w_1}{\partial q_j} \right) \right]_0^\infty
  - \left[ \frac{\partial w_1}{\partial z} \frac{\partial \varepsilon}{\partial x_j} \frac{\partial w_1}{\partial q_j} \right]_0^\infty
  - \left[ \varepsilon \frac{\partial}{\partial z} \bigg( \frac{\partial w_1}{\partial x_j} \bigg) \frac{\partial w_1}{\partial q_j} \right]_0^\infty \nonumber \\
  &\hspace*{3.9cm}
  + \int_0^\infty \varepsilon \frac{\partial}{\partial z} \bigg( \frac{\partial w_1}{\partial x_j} \bigg) \frac{\partial}{\partial z} \bigg( \frac{\partial w_1}{\partial q_j} \bigg) d z'
  + \int_0^\infty \frac{\partial w_1}{\partial x_j} \varepsilon \frac{q^2}{\hbar^2} \frac{\partial w_1}{\partial q_j} d z' .
  \label{eq:upperhalfinhomogeneous-second-term}
\end{align}
Subtracting Eq.~(\ref{eq:upperhalfinhomogeneous-third-term}) from Eq.~(\ref{eq:upperhalfinhomogeneous-second-term}), we see that the remaining integrals cancel, and we obtain Eq.~(\ref{eq:upperhalfinhomogeneous-middle-terms}).

Inserting the results~(\ref{eq:upperhalfinhomogeneous-first-term}), (\ref{eq:upperhalfinhomogeneous-last-term}) and~(\ref{eq:upperhalfinhomogeneous-middle-terms}) in Eq.~(\ref{eq:upperhalfinhomogeneous}) and evaluating all boundary terms, we obtain
\begin{align}
  \int_0^{\infty} g(0,z') f_{1s}(z') dz' 
    &= -\frac{i}{2} \frac{2 \pi e^2 \hbar}{\varepsilon_{\mathrm{eff}}|\mathbf{q}| }\frac{\Pi_0}{w_1} \frac{w_2}{\det(W)} \sum_j \frac{\partial}{\partial x_j} \left(  w_1 \varepsilon \frac{\partial}{\partial z} \left(\frac{\partial w_1}{ \partial q_j}\right)- \frac{\partial w_1}{\partial q_j} \varepsilon \frac{\partial w_1}{\partial z}\right)\nonumber\\
    &\qquad + i \sum_j \frac{\partial }{\partial x_j}\left(\frac{2 \pi e^2 \hbar}{\varepsilon_{\mathrm{eff}}|\mathbf{q}|} \frac{\Pi_0}{w_1}\right) \frac{w_2}{\det(W)} \left(- w_1 \varepsilon \frac{\partial }{\partial z} \left(\frac{\partial w_1}{\partial q_j} \right) +  \frac{\partial w_1}{\partial z} \varepsilon  \frac{\partial w_1}{\partial q_j}\right) ,
\end{align}
where all functions of $z$ are to be evaluated at $z=0$ from here on.
We can then use the relation
\begin{equation}
  w_i^2\frac{\partial}{\partial q_j} \left(\frac{1}{w_i} \varepsilon \frac{\partial w_i}{\partial z}\right) = w_i \varepsilon  \frac{\partial }{\partial z}  \left(\frac{\partial w_i }{\partial q_j} \right) -\frac{\partial w_i}{\partial q_j} \varepsilon  \frac{\partial w_i}{\partial z},
\end{equation}
to find
\begin{align}
  \int_0^{\infty} g(0,z') f_{1s}(z')dz' &= -\frac{i}{2} \frac{2 \pi e^2 \hbar}{\varepsilon_{\mathrm{eff}}|\mathbf{q}| }\frac{\Pi_0}{w_1} \frac{w_2}{\det(W)} \sum_j \frac{\partial}{\partial x_j} \left( w_1^2\frac{\partial}{\partial q_j} \left(\frac{1}{w_1} \varepsilon \frac{\partial w_1}{\partial z}\right)\right)\nonumber\\
  &\qquad - i \sum_j \frac{\partial }{\partial x_j}\left(\frac{2 \pi e^2 \hbar}{\varepsilon_{\mathrm{eff}}|\mathbf{q}|} \frac{\Pi_0}{w_1}\right) \frac{w_2}{\det(W)} w_1^2\frac{\partial}{\partial q_j} \left(\frac{1}{w_1} \varepsilon \frac{\partial w_1}{\partial z}\right).
\end{align}
Performing the derivatives of $w_1$ with respect to $x_j$, this can also be written as
\begin{align}
  \int_0^{\infty} g(0,z') f_{1s}(z')dz' 
  &= -\frac{i}{2} \frac{2 \pi e^2 \hbar}{\varepsilon_{\mathrm{eff}}|\mathbf{q}| }\Pi_0 \frac{w_1 w_2}{\det(W)} \sum_j \frac{\partial^2}{\partial x_j \partial q_j} \left(\frac{1}{w_1} \varepsilon \frac{\partial w_1}{\partial z}\right)\nonumber\\
  &\qquad - i \sum_j \frac{\partial }{\partial x_j}\left(\frac{2 \pi e^2 \hbar}{\varepsilon_{\mathrm{eff}}|\mathbf{q}|} \Pi_0\right) \frac{w_1 w_2}{\det(W)} \frac{\partial}{\partial q_j} \left(\frac{1}{w_1} \varepsilon \frac{\partial w_1}{\partial z}\right) .
  \label{eq:upperhalf-integral-outcome}
\end{align}

The integral over the interval $(-\infty,0]$ can be performed in exactly the same way as in Eqs.~(\ref{eq:integration-by-parts-ex}),~(\ref{eq:upperhalfinhomogeneous-last-term}) and~(\ref{eq:upperhalfinhomogeneous-middle-terms}). When one interchanges $w_1$ and $w_2$ in those outcomes, and changes the integration limits from  $[0,\infty)$ to $(-\infty,0]$, one obtains the results for the interval $(-\infty,0]$. One can then perform the same steps, to arrive at a result similar to Eq.~(\ref{eq:upperhalf-integral-outcome}). Note, however, that the change of integration limits leads to a relative minus sign between the upper and lower half, since for the lower half all boundary terms vanish at the lower limit of integration.

Combining both results, we find that the integral in Eq.~(\ref{eq:gamma1forz0}) equals
\begin{align}
  \int_{-\infty}^{\infty} g(0,z') f_{1s}(z')dz' 
  &= \frac{i}{2} \frac{2 \pi e^2 \hbar}{\varepsilon_{\mathrm{eff}}|\mathbf{q}| }\Pi_0 \frac{w_1 w_2}{\det(W)} \sum_j \frac{\partial^2}{\partial x_j \partial q_j} \left(\frac{1}{w_2} \varepsilon \frac{\partial w_2}{\partial z} - \frac{1}{w_1} \varepsilon \frac{\partial w_1}{\partial z}\right)\nonumber\\
  &\qquad + i \sum_j \frac{\partial }{\partial x_j}\left(\frac{2 \pi e^2 \hbar}{\varepsilon_{\mathrm{eff}}|\mathbf{q}|} \Pi_0\right) \frac{w_1 w_2}{\det(W)} \frac{\partial}{\partial q_j} \left(\frac{1}{w_2} \varepsilon \frac{\partial w_2}{\partial z} - \frac{1}{w_1} \varepsilon \frac{\partial w_1}{\partial z}\right) .
  \label{eq:total-integral-outcome}
\end{align}
We can now use Eqs.(\ref{eq:definition-effective-dielectricconstant}) and~((\ref{eq:definition-effective-dielectricconstant2})), which show that $w_1 w_2/\det(W) = \hbar/(2 |\mathbf{q}| \varepsilon_{\mathrm{eff}})$. This implies that
\begin{align}
  \int_{-\infty}^{\infty} g(0,z') f_{1s}(z')dz' 
    = \frac{i}{2} \frac{2 \pi e^2 \hbar}{(\varepsilon_{\mathrm{eff}}|\mathbf{q}|)^2 }\Pi_0\sum_j \frac{\partial^2 (\varepsilon_{\mathrm{eff}} |\mathbf{q}|)}{\partial x_j \partial q_j}  
    + \frac{i}{\varepsilon_{\mathrm{eff}} |\mathbf{q}|}  \sum_j  \frac{\partial }{\partial x_j}\left(\frac{2 \pi e^2 \hbar}{\varepsilon_{\mathrm{eff}} |\mathbf{q}|} \Pi_0 \right) \frac{\partial (\varepsilon_{\mathrm{eff}} |\mathbf{q}|)}{\partial q_j} .
\end{align}
This leads us to our final result for $\Gamma_1$ at $z=0$, Eq.~(\ref{eq:gamma1forz0}). Using Eq.~(\ref{eq:generalsol-Green}) to compute $g(0,0)$, we find
\begin{equation}
  \Gamma_1(z=0) 
    = \frac{2 \pi e^2 \hbar}{\varepsilon_{\mathrm{eff}}|\mathbf{q}| }\Pi_1 
    + \frac{i}{2} \frac{2 \pi e^2 \hbar}{(\varepsilon_{\mathrm{eff}}|\mathbf{q}|)^2 }\Pi_0\sum_j \frac{\partial^2 (\varepsilon_{\mathrm{eff}} |\mathbf{q}|)}{\partial x_j \partial q_j}  
    + i \sum_j  \frac{\partial }{\partial x_j}\left(\frac{2 \pi e^2 \hbar}{\varepsilon_{\mathrm{eff}} |\mathbf{q}|} \Pi_0 \right) \frac{1}{\varepsilon_{\mathrm{eff}} |\mathbf{q}|} \frac{\partial (\varepsilon_{\mathrm{eff}} |\mathbf{q}|)}{\partial q_j} .
  \label{eq:Gamma1-z0-final}
\end{equation}

As discussed in Sec.~\ref{subsec:reviewDerivAmplitude}, it is the quantity
\begin{equation}
  \mathcal{H}_1+ \frac{i}{2}\frac{\partial^2 \mathcal{H}_0}{\partial x_j \partial q_j},
\end{equation}
that enters in the amplitude of the potential, where $\mathcal{H}_1 = -\Gamma_1(z=0)$.
In the final part of this appendix, we compute this quantity, and show that it can be cast in the form given in Eq.~(\ref{eq:AmplitudeExponent}). Since $\mathcal{H}_0$ is given by Eq.~(\ref{eq:effective-classical-Hamiltonian}), we have
\begin{align}
  \frac{\partial^2 \mathcal{H}_0}{\partial x_j \partial q_j} 
  &= \frac{2 \pi e^2 \hbar}{(\varepsilon_{\mathrm{eff}}|\mathbf{q}|)^2 }\Pi_0 \frac{\partial^2 (\varepsilon_{\mathrm{eff}}|\mathbf{q}|)}{\partial x_j \partial q_j} 
  + \frac{\partial }{\partial x_j}\left(\frac{2 \pi e^2 \hbar}{\varepsilon_{\mathrm{eff}}|\mathbf{q}|} \Pi_0 \right) \frac{1}{ \varepsilon_{\mathrm{eff}} |\mathbf{q}| } \frac{\partial (  \varepsilon_{\mathrm{eff}} |\mathbf{q}| )}{\partial q_j} \nonumber \\
  &\qquad + \frac{\partial }{\partial q_j}\left(\frac{2 \pi e^2 \hbar}{\varepsilon_{\mathrm{eff}}|\mathbf{q}|} \Pi_0 \right) \frac{1}{ \varepsilon_{\mathrm{eff}} |\mathbf{q}|} \frac{\partial ( \varepsilon_{\mathrm{eff}} |\mathbf{q}|)}{\partial x_j} 
  - \frac{2 \pi e^2 \hbar}{\varepsilon_{\mathrm{eff}}|\mathbf{q}| } \frac{\partial^2 \Pi_0}{\partial x_j \partial q_j}.
  \label{eq:H0-second-deriv}
\end{align}
Combining Eqs.(\ref{eq:Gamma1-z0-final}) and~(\ref{eq:H0-second-deriv}), we find
\begin{align}
  \mathcal{H}_1 + \frac{i}{2} \sum_j \frac{\partial^2 \mathcal{H}_0}{\partial x_j \partial q_j} 
  &= -\frac{2 \pi e^2 \hbar}{\varepsilon_{\mathrm{eff}}|\mathbf{q}| } \bigg(\Pi_1 + \frac{i}{2}\sum_j \frac{\partial^2 \Pi_0}{\partial x_j \partial q_j} \bigg) 
  + \frac{i}{2} \sum_j  \frac{\partial }{\partial q_j}\left(\frac{2 \pi e^2 \hbar}{\varepsilon_{\mathrm{eff}}|\mathbf{q}|} \Pi_0 \right) \frac{\partial \ln ( \varepsilon_{\mathrm{eff}} |\mathbf{q}|)}{\partial x_j} \nonumber \\
  & \qquad- \frac{i}{2}  \sum_j  \frac{\partial }{\partial x_j}\left(\frac{2 \pi e^2 \hbar}{\varepsilon_{\mathrm{eff}}|\mathbf{q}|} \Pi_0 \right) \frac{\partial \ln ( \varepsilon_{\mathrm{eff}} |\mathbf{q}|)}{\partial q_j}.
\end{align}
Using the definition of the Poisson bracket
\begin{equation}  \label{eq:def-Poisson-bracket}
  \{f,g\} = \sum_j \left(\frac{\partial f}{\partial x_j}\frac{\partial g}{\partial q_j}- \frac{\partial f}{\partial q_j}\frac{\partial g}{\partial x_j}\right),
\end{equation}
and the definition of $\mathcal{H}_0$, see Eq.~(\ref{eq:effective-classical-Hamiltonian}), we finally arrive at Eq.~(\ref{eq:AmplitudeExponent}).

\subsection{Derivation of an expression for the energy density}
\label{subapp:energydensity}

In this appendix, we compute the integrated energy density for the potential $V(\mathbf{x},z)$ given by Eq.~(\ref{eq:full-potential}). We integrate expression~(\ref{eq:totalenergydensity}), which was discussed in Ref.~\cite{Koskamp23}, and show that Eq.~(\ref{eq:inplaneenergydensity}) holds.

We first note that $\nabla$ in Eq.~(\ref{eq:totalenergydensity}) is the three-dimensional gradient, that is, $\nabla = \left(\partial/\partial \mathbf{x},\partial/\partial z\right)$.
Without much loss of generality, we may assume that $\Gamma_0$ is real. This roughly corresponds to an Hermitian Hamiltonian, cf. Eq.~(\ref{eq:defEffClasHam}). In the terminology introduced in Sec.~\ref{sec:bound-states}, see also Refs.~\cite{Reijnders22,Koskamp23}, we may say that this situation corresponds to plasmons in a classically allowed region. Note that having a real-valued $\Gamma_{0}$ corresponds to having real-valued functions $w_i$, see Eq.~(\ref{eq:definition-principal-symbol}).
Substituting the potential~(\ref{eq:full-potential}) in Eq.~(\ref{eq:totalenergydensity}) and only taking the leading-order terms in $\hbar$ into account, we find
\begin{equation}
  |\nabla V|^2 = \left|\frac{\partial V}{\partial \mathbf{x}}\right|^2 + \left|\frac{\partial V}{\partial z}\right|^2 = \frac{|A_0^0|^2}{|J(\mathbf{x})|} \frac{1}{\varepsilon_{\mathrm{eff}} |\partial S / \partial \mathbf{x}|} \left(\Gamma_0(z) \frac{1}{\hbar^2}\left|\frac{\partial S}{\partial \mathbf{x}}\right|^2 \Gamma_0(z) + \frac{\partial \Gamma_0}{\partial z}\frac{\partial \Gamma_0}{\partial z} \right).
\end{equation}
We note that taking the derivative of either $\Gamma_0$ or the amplitude with respect to $\mathbf{x}$ leads to higher-order terms in $\hbar$, which we therefore neglect.
Moreover, we remark that this result holds regardless of the Berry phase in Eq.~(\ref{eq:full-potential}), since it cancels upon taking the absolute value.

We consider the integrated energy density $\mathcal{U}_\mathrm{I}(\mathbf{x})$, defined as the integral of the energy density $U(\mathbf{x},z)$ over $z$, that is  
\begin{equation}
  \mathcal{U}_\mathrm{I}(\mathbf{x})
    = \int_{-\infty}^\infty \mathcal{U}(\mathbf{x},z) dz 
    = \frac{1}{16 \pi e^2 } \frac{|A_0^0|^2}{|J(\mathbf{x})|} \frac{1}{\varepsilon_{\mathrm{eff}} |\partial S / \partial \mathbf{x}|} \int_{-\infty}^{\infty} \left(\varepsilon (\mathbf{x},z) \Gamma_0(z) \frac{1}{\hbar^2}\left|\frac{\partial S}{\partial \mathbf{x}}\right|^2 \Gamma_0(z) + \varepsilon (\mathbf{x},z)\frac{\partial \Gamma_0}{\partial z}\frac{\partial \Gamma_0}{\partial z} \right) dz.
\end{equation}
As in the previous appendix, we separate the integral into two parts, corresponding to the upper $[0,\infty)$, and lower $(-\infty,0]$ halves of the system. Since $\Gamma_0$ is symmetric in $z=0$ upon interchanging $w_1$ and $w_2$, see Eq.~(\ref{eq:definition-principal-symbol}), one can infer the outcome for the lower half from the outcome for the upper half. Considering the upper half, and inserting our expression~(\ref{eq:definition-principal-symbol}) for $\Gamma_0$, we obtain
\begin{equation}
  \int_0^{\infty} \mathcal{U} dz =  \frac{1}{16 \pi e^2 } \frac{|A_0^0|^2}{|J(\mathbf{x})|} \frac{1}{\varepsilon_{\mathrm{eff}} |\partial S / \partial \mathbf{x}|} \int_0^{\infty} \frac{1}{w^2_1(0)} \left(w_1(z)  \frac{\varepsilon (\mathbf{x},z)}{\hbar^2} \left|\frac{\partial S}{\partial \mathbf{x}}\right|^2 w_1(z) +\frac{\partial w_1}{\partial z} \varepsilon (\mathbf{x},z)\frac{\partial w_1 }{\partial z} \right) dz ,
\end{equation} 
where we used the Hamilton-Jacobi equation $\mathcal{H}_0 (\mathbf{x},\partial S/ \partial \mathbf{x}) = 0$ to set $2 \pi e^2 \hbar \Pi_0/(\varepsilon_\mathrm{eff} |\partial S/\partial \mathbf{x}|) = 1$.
Integrating the second term by parts, we find
\begin{align}
  \int_0^{\infty} \mathcal{U} dz 
  &= \frac{1}{16 \pi e^2 }  \frac{|A_0^0|^2}{|J(\mathbf{x})|} \frac{1}{\varepsilon_{\mathrm{eff}} |\partial S / \partial \mathbf{x}|} \frac{1}{w^2_1(0)} \Bigg(\int_0^{\infty}w_1(z) \bigg(\frac{\varepsilon (\mathbf{x},z)}{\hbar^2} \left|\frac{\partial S}{\partial \mathbf{x}}\right|^2 w_1(z) - \frac{\partial}{\partial z} \bigg(\varepsilon (\mathbf{x},z)\frac{\partial w_1 }{\partial z}\bigg)\bigg) dz   \nonumber \\  
  & \qquad + \left[w_1(z) \varepsilon (\mathbf{x},z)\frac{\partial w_1 }{\partial z}\right]_0^{\infty}  \Bigg).
\end{align} 
The remaining integral on the right-hand side vanishes because $w_1(z)$ satisfies the homogeneous differential equation $\hat{F}_0 w_1(z) = 0$, cf. the discussion above Eq.~(\ref{eq:relationF1F0afterproduct}) in the previous appendix.

Since both $w_1$ and its derivatives go to zero as $z \to \infty$, part of the boundary term also vanishes, and we are left with the contribution at $z=0$.
Adding the contribution from the lower half, which comes with a relative minus sign because the boundaries are $-\infty$ and $0$, we obtain 
\begin{align}
  \mathcal{U}_\mathrm{I}(\mathbf{x})
    = \int_{-\infty}^{\infty} \mathcal{U}(\mathbf{x},z) dz 
    = \frac{1}{16 \pi e^2 }  \frac{|A_0^0|^2}{|J(\mathbf{x})|} \frac{1}{\varepsilon_{\mathrm{eff}} |\partial S / \partial \mathbf{x}|} \left( -\frac{1}{w_1(0)} \varepsilon(\mathbf{x},0)\frac{\partial w_1(0) }{\partial z} + \frac{1}{w_2(0)} \varepsilon(\mathbf{x},0)\frac{\partial w_2 (0)}{\partial z} \right).
\end{align}  
Finally, using the definition of $\varepsilon_{\mathrm{eff}}$ from Eq.~(\ref{eq:definition-effective-dielectricconstant2}), we find  
\begin{equation}
  \mathcal{U}_\mathrm{I}(\mathbf{x})
  = \frac{1}{16 \pi e^2 } \frac{|A_0^0|^2}{|J(\mathbf{x})|} \frac{2 \varepsilon_{\mathrm{eff}} |\partial S / \partial \mathbf{x}|}{\hbar \varepsilon_{\mathrm{eff}} |\partial S / \partial \mathbf{x}|}  
  = \frac{1}{8 \pi e^2 \hbar } \frac{|A_0^0|^2}{|J(\mathbf{x})|} .
\end{equation}
This result exactly coincides with Eq.~(81) in Ref.~\cite{Koskamp23}. However, this time we started from an arbitrary model for $\varepsilon(\mathbf{x},z)$, instead of a simplified model.

\end{widetext}

\section{Simple turning point}\label{ap:SimpleTurningPoint}
The behavior of $q_x$ around a turning point, where it vanishes, is fundamental to understanding the nature of plasmonic bound states. In this appendix, we demonstrate that these turning points are simple turning points~\cite{Maslov81,Arnold85,Poston78,Berry72}, independent of the system parameters.

To analyze this behavior, we must consider the role of the effective dielectric function $\varepsilon_{\mathrm{eff}}(\mathbf{x},\mathbf{q})$. While $\varepsilon_{\mathrm{eff}}(\mathbf{x},\mathbf{q})$ is momentum dependent, we argue that this variation does not qualitatively alter the scaling of $q_x$ near the turning point. To justify this statement, let us examine two limiting cases: as seen in Fig.~\ref{fig:BackgroundDielectricOverXandQ}(b), in the large $\mathbf{|q|}$ limit, screening becomes constant at $\varepsilon_{\mathrm{M}}$. Conversely, in the small $\mathbf{|q|}$ limit, the system behaves similarly as previously analyzed in Ref.~\cite{Koskamp23}. At both these limits, there are no discontinuities or divergences in $\varepsilon_{\mathrm{eff}}(\mathbf{x},\mathbf{q})$. Moreover, it behaves smoothly as a function of $\mathbf{q}$, meaning that we can perform a Taylor expansion to first order in $\mathbf{q}$. We therefore argue that the variation of $\varepsilon_{\mathrm{eff}}(\mathbf{x},\mathbf{q})$ does not modify the behavior of $q_x$ at the turning points.

When $q_y$ is small, we can analyze the behavior near the turning points analytically. We start our analysis from the effective classical Hamiltonian, Eq.~(\ref{eq:effective-classical-Hamiltonian}), which in the limit of small momenta $|\mathbf{q}| \ll 1$ takes the approximate form  
\begin{equation}
  \mathcal{H}_0 \approx 1- \frac{g_\mathrm{s} e^2 p_\mathrm{F}^2(\mathbf{x}) |\mathbf{q}|}{2 m \varepsilon_\mathrm{avg}(\mathbf{x}) \hbar E^2} + \mathcal{O}(|\mathbf{q}^2|),
\end{equation}
where the average dielectric constant is defined as  
\begin{equation}
  2 \varepsilon_\mathrm{avg} = \varepsilon_{\mathrm{A}} + \varepsilon_{\mathrm{B}},
\end{equation}
which follows naturally from the effective dielectric function $\varepsilon_{\mathrm{eff}}(|\mathbf{q}| \to 0)$. Since plasmons are defined by the equation $\mathcal{H}_0 = 0$, we directly have  
\begin{equation}
  |\mathbf{q}|  = \frac{2 m \varepsilon_\mathrm{avg}(x) \hbar E^2}{g_\mathrm{s} e^2 p_\mathrm{F}^2(x)}.
\end{equation}
Expressing the total momentum in terms of its components, $|\mathbf{q}|^2 = q_x^2 + q_y^2$, we obtain  
\begin{equation}
  q^2_x  = \frac{2 m \varepsilon_\mathrm{avg}(x) \hbar E^2}{g_\mathrm{s} e^2 p_\mathrm{F}^2(x)} -  q_y^2.
\end{equation}

By definition, $q_x=0$ at the turning point, which gives us a relation between $x_\mathrm{c}$, $q_y$ and $E$.
\begin{equation}\label{eq:qyresultxc}
  q^2_y  =  \left(\frac{2 m \varepsilon_\mathrm{avg}(x_\mathrm{c}) \hbar E^2}{g_\mathrm{s} e^2 p_\mathrm{F}^2(x_\mathrm{c})}\right)^2.
\end{equation}
To determine the nature of the turning point, we Taylor-expand $q_x^2$ around $x_\mathrm{c}$, which yields
\begin{align}
  q_x^2 &= \left(\frac{2 m \varepsilon_\mathrm{avg}(x_\mathrm{c}) \hbar E^2}{g_\mathrm{s} e^2 p_\mathrm{F}^2(x_\mathrm{c})}\right)^2  - q_y^2 \nonumber\\
  &\qquad + (x-x_\mathrm{c}) \left(\frac{2 m \varepsilon_\mathrm{avg}(x) \hbar E^2}{g_\mathrm{s} e^2 p_\mathrm{F}^2(x)}\right)^2\nonumber\\
   & \qquad \qquad \times 
   \left( 2 \frac{ \varepsilon'_{\mathrm{avg}}(x)}{\varepsilon_{\mathrm{avg}}(x)} \right. 
   \left. \left. - 4 \frac{  p_\mathrm{F}'(x) }{p_\mathrm{F}(x)}\right)\right|_{x_\mathrm{c}}.
\end{align}
With Eq.~(\ref{eq:qyresultxc}), we can substitute $q_y$ as a function of $x_\mathrm{c}$ and $E$. We find
\begin{align}
  q_x^2 &= (x-x_\mathrm{c}) \left(\frac{2 m \varepsilon_\mathrm{avg}(x) \hbar E^2}{g_\mathrm{s} e^2 p_\mathrm{F}^2(x)}\right)^2 \left( 2 \frac{ \varepsilon'_{\mathrm{avg}}(x)}{\varepsilon_{\mathrm{avg}}(x)} \right. \nonumber\\
  & \qquad \qquad \left. \left. - 4 \frac{  p_\mathrm{F}'(x) }{p_\mathrm{F}(x)}\right)\right|_{x_\mathrm{c}}.
\end{align}
By definition, this confirms that the turning point is simple, as $q_x^2$ depends linearly on $(x - x_\mathrm{c})$. To eliminate the explicit dependence on $x_\mathrm{c}$, we may express $x_\mathrm{c}$ as a function of $(q_y, E)$, i.e., $x_\mathrm{c} = x_\mathrm{c}(q_y, E)$. 

For small $q_y$, we have thus explicitly shown that the turning points are simple. For larger values of $q_y$, we numerically confirmed that the proportionality $q_x^2 \propto x-x_c$ continues to hold, meaning that the turning points remain simple.

\end{document}